\documentclass[aps,floats]{revtex4-2}
\usepackage{amsmath,amssymb}
\usepackage{graphicx,epsfig}
\usepackage[greek,english]{babel}
\usepackage{bbold}

\numberwithin{equation}{section}
 
\begin{document}
\bibliographystyle {plain}

\pdfoutput=1
\def\oppropto{\mathop{\propto}} 
\def\opsimeq{\mathop{\simeq}}
\def\opoverderline{\mathop{\overline}}
\def\operarrow{\mathop{\longrightarrow}}
\def\opsim{\mathop{\sim}}

\def\opmin{\mathop{\min}} 
\def\opmax{\mathop{\max}} 
\def\oplim{\mathop{\lim}}

\title{ Real-Space-RG for the drawups and drawdowns of run-and-tumble trajectories: \\  
universality classes for models with space-dependent velocities and switching rates   } 


\author{C\'ecile Monthus}
\affiliation{Universit\'e Paris-Saclay, CNRS, CEA, Institut de Physique Th\'eorique, 91191 Gif-sur-Yvette, France}


\begin{abstract}
For one-dimensional run-and-tumble processes with the two velocities $v_{\pm}(x)$ and the two switching rates $\gamma_{\pm}(x)$, a given stochastic trajectory is an alternating sequence of local minima and local maxima corresponding to the switching events between the  ballistic motions with velocities $v_{\pm}(x) $ in the two directions. The renormalized trajectory at scale $\Gamma$ is obtained by keeping only the alternating sequence of local minima and local maxima that are separated by drawups and drawdowns of amplitude bigger than $\Gamma$. Its statistical properties can be analyzed via the generalization of the previous work arxiv:2611.04853 concerning diffusion processes involving arbitrary diffusion coefficients $D(x)$ and Ito forces $F_I(x)$. The main result is that the scale function $s(x)$ of diffusion processes that depends only on the ratio $\frac{F_I(x)}{D(x) } $ is replaced by a pair of scale functions $s_{\pm}(x)$ that depend only on the two ratios $\frac{\gamma_+( x ) }{v_+( x )}$ and $\frac{\gamma_-( x ) }{v_-( x )}$. This pair of scale functions $s_{\pm}(x)$ govern the exit probabilities from an interval and determine the universality classes of the renormalized trajectories: all run-and-tumble processes with the same two ratios $\frac{\gamma_+( x ) }{v_+( x )}$ and $\frac{\gamma_-( x ) }{v_-( x )}$ share the same statistics of renormalized trajectories. This general framework is illustrated with examples that are compared to their diffusive counterparts : uniform symmetric and asymmetric processes on the whole line $x \in]-\infty,+\infty[$, geometric process on the half-line $x \in]0,+\infty[$, and an exponential functional of the telegraph process with a normalizable steady state.

\end{abstract}

\maketitle

\section{ Introduction }

For Brownian trajectories with or without drift, the idea to analyze the renormalized trajectories at scale $\Gamma$, where one keeps only the alternating sequence of local minima and local maxima that are separated by drawups and drawdowns of amplitude bigger than $\Gamma$,
has appeared in the context of the Strong-Disorder-RG approach (see the reviews \cite{igloi05review,igloiMiniReview} and references therein)
where the goal was to analyze the ground states and low-energy properties of random quantum spin chains,  
either with antiferromagnetic couplings \cite{ma79,danielantiferro}
or with random transverse fields \cite{fisher92,rgfinitesize,danielreview}.
Indeed for the Random Transverse Field Ising Chain, the logarithms of the random transverse fields and of the random couplings build a one-dimensional 
random walk whose drift vanishes at the critical point
\cite{igloi98,dharyoung,microcano}, and the Strong-Disorder-RG rules amount to eliminate
iteratively the drawups and drawdowns of the smallest amplitude $\Gamma$ representing the RG scale. As a consequence, exactly the same RG-rules have been used to 
analyze the anomalously slow dynamics of random walks in Brownian landscapes
where the particle has to overcome larger and larger barriers
\cite{igloi98c,rgsinaishort,rgsinailong,rgrfim,rgreadiff,locgolosov,energysinai}.
However, the Strong-Disorder RG approach that becomes asymptotically exact at large scale $\Gamma$ for these various disordered models is actually an exact procedure to compute the statistics of local extrema of the Brownian motion
as a function of the scale $\Gamma$. It is thus interesting to apply this renormalization
procedure based on the drawups and drawdowns to other stochastic trajectories,
and explicit results have been obtained for diffusions with the uniform diffusion coefficient $D(x)=1$ \cite{rgtoy},
as well as for the Brownian motion with various constraints, for Bessel processes and for some continuous time random walks in \cite{schehr10}. 
Note that in the mathematical literature, the notion of $\Gamma$-extrema has been also analyzed under the name h-extrema \cite{Neveu,Foggionato}, while 
the statistics of the maximal drawdown
has been much studied for various processes (see \cite{Taylor75,Lehoczky77,2009,Z2012,Z2015,proof,Salminen,BookZhang} and references therein),
in particular in relation with financial applications.

In the recent work \cite{c_RGdiff}, the explicit solutions for the RG-flows 
have been obtained for any one-dimensional diffusion process
involving an arbitrary diffusion coefficient $D(x)$ and an arbitrary Ito force $F_I(x)$,
as well as for discrete-time and continuous-time random walks on the one-dimensional lattice with arbitrary nearest-neighbor jump probabilities or rates.
The goal of the present paper is to adapt this framework to any 
one-dimensional run-and-tumble process involving the two velocities $v_{\pm}(x)$ and the two switching rates $\gamma_{\pm}(x)$, since run-and-tumble particles have attracted
a lot of interest recently in statistical physics
\cite{tailleur2008,angelani2014,angelani2015,evans2016,evans2017,redner2018,satya2018,evans2018,emil2018,pld2019,harris,emil2019,kundu2019,angelani,emil2020,satya2019,hartmann2020,satya2020,rosso2020,gregory2019,kundu2020,pld2020,entropyprod,bressloff2020,localtime,gregory2021a,boyer2021,bressloff2021,gregory2021b,c_largeDevRun,Mori,Kundu,Naftali,FvW,mathis24,Leo1,Leo2,mathis25,bressloff25,Basu,mathis26}.

The paper is organized as follows.
In Section \ref{sec_run}, we introduce the run-and-tumble model
and write the exit probabilities out of an interval in terms of the pair of scale functions $s_{\pm}(x)$ that will play a major role in the whole paper.
In Section \ref{sec_bond}, we describe how the bond probabilities of the renormalized trajectories at scale $\Gamma$ can be computed via a direct probability analysis
and write the corresponding RG-flows as a function of the RG-scale $\Gamma$.
In Section \ref{sec_nextrema}, we analyze the corresponding 
densities $n_{\Gamma}^{\pm}$ of maxima and minima in the renormalized trajectories.
This general framework is illustrated with various examples in section \ref{sec_UniversalityClasses}.
Our conclusions are summarized in section \ref{sec_conclusion}.
In the Appendices, we take into account the additional durations of the drawups and drawdowns of the renormalized trajectories, in order to
obtain explicit results for the time-Laplace transforms 
as a function of the Laplace variable $p>0$ (instead of $p=0$ in the main text,
where one keeps only the values of the local maxima and minima).


\section{ Run-and-tumble process with velocities $v_{\pm}(x) >0$ and switching rates $ \gamma_{\pm}(x)> 0$ }

\label{sec_run}

The particle is characterized by its position $x(t) \in ]-\infty,+\infty[$ and by its internal state $\sigma(t)=\pm$.
The dynamical rules involve two space-dependent velocities $v_{\pm}(x) >0$ and two space-dependent switching rates $ \gamma_{\pm}(x)>0 $ as follows:

(i) the position $x(t)$ moves ballistically with the positive velocity $v_+(x)>0$ if $\sigma(t)=+1$ and with the negative velocity $(-v_-(x))<0$ if $\sigma(t)=-1$
\begin{eqnarray}
\frac{dx(t)}{dt} = \sigma(t) v_{\sigma(t)} (x(t))
\label{runx}
\end{eqnarray}

(ii) the internal state $\sigma(t)$ can switch towards the opposite internal state 
$[-\sigma(t)]$ with the rate $\gamma_{\sigma(t)}(x(t)) \geq 0 $
\begin{eqnarray}
\sigma(t+dt) = 
\begin{cases}
- \sigma(t) \ \ \ \text{with probability  } dt \gamma_{\sigma(t)} (x(t)) \\
\sigma(t)  \ \ \ \text{with probability  } [1- dt \gamma_{\sigma(t)} (x(t)) ]
\end{cases}
\label{runsigma}
\end{eqnarray}

\subsection{ The generator ${\cal L} $ governing the dynamics of the probabilities $P_t(x,\sigma=\pm)$ }
 
The dynamical system satisfied by the probabilities $P_t( x,\pm)   $ to be at position $x$ and in the internal state $\sigma=\pm $ at time $t$
\begin{eqnarray}
 \partial_t P_t(x,+)   &&   =  -    \partial_x  \left[ v_+( x ) P_t(x,+)  \right]    -   \gamma_+( x ) P_t(x,+)   +   \gamma_-( x ) P_t(x,-) 
\nonumber \\
 \partial_t P_t(x,-)  &&   =   -  \partial_x   \left[  -  v_-( x ) P_t(x,-)  \right]    +   \gamma_+( x ) P_t(x,+)   -   \gamma_-( x ) P_t(x,-) 
\label{run}
\end{eqnarray}
can be rewritten in the matrix form
\begin{eqnarray}
 \partial_t
\begin{pmatrix} 
 P_t(x,+)
 \\ P_t(x,-)
  \end{pmatrix}
= \begin{pmatrix} 
-   \gamma_+( x ) -    \partial_x   v_+( x )         &     \gamma_-( x )   \\
\gamma_+( x )  & -   \gamma_-( x )  + \partial_x       v_-( x )    
 \end{pmatrix} 
  \begin{pmatrix} 
 P_t(x,+)
 \\  P_t(x,-) 
  \end{pmatrix}
  \equiv
     {\cal L}
  \begin{pmatrix} 
 P_t(x,+)
 \\  P_t(x,-) 
  \end{pmatrix}
\label{rungenerator}
\end{eqnarray}
that involves the $2 \times 2$ first-order-differential generator ${\cal L} $.

The sum of the two equations of Eq. \ref{run}
yields that the dynamics of the total probability $P_t(x) \equiv  P_t(x,+) + P_t(x,-)$ at position $x$
\begin{eqnarray}
 \partial_t P_t(x) \equiv  \partial_t \left(P_t(x,+) + P_t(x,-)\right)  &&   =  -    \partial_x  \left[ v_+( x ) P_t(x,+)  -  v_-( x ) P_t(x,-)\right]    
 \equiv -    \partial_x J_t(x)
\label{runcontinuity}
\end{eqnarray}
is a continuity equation involving the current $J_t(x)$ at position $x$
\begin{eqnarray}
J_t(x) \equiv v_+( x ) P_t(x,+)  -  v_-( x ) P_t(x,-) \equiv J_t^{+}(x)-J_t^{-}(x)
\label{runcurrent}
\end{eqnarray}
obtained from the difference between the two ballistic currents $J_t^{\sigma=\pm}(x) 
\equiv v_{\sigma}( x ) P_t(x,\sigma)$ associated to the internal states $\sigma=\pm$.


\subsection{ The adjoint ${\cal L}^{\dagger} $ governing the dynamics of averaged values of observables}

The dynamics of the averaged values ${\mathbb E}_t [ O(x,\sigma ) ] $ of observables $O(x,\sigma)$ at time $t$
\begin{eqnarray}
 {\mathbb E}_t [ O(x,\sigma )  ] \equiv \sum_{\sigma=\pm} \int dx O(x,\sigma) P_t(x,\sigma)
\label{runobs}
\end{eqnarray}
can be obtained from Eq. \ref{rungenerator}
\begin{eqnarray}
 \partial_t   {\mathbb E}_t [ O(x,\sigma )  ] && =  \sum_{\sigma=\pm} \int dx O(x,\sigma)  \partial_tP_t(x,\sigma)
 \nonumber \\
 && = \sum_{\sigma=\pm} \int dx O(x,\sigma)
 \bigg(  -    \partial_x  \left[ \sigma v_{\sigma}( x ) P_t(x,\sigma)  \right]    -   \gamma_{\sigma}( x ) P_t(x,\sigma)   +   \gamma_{-\sigma}( x ) P_t(x,-\sigma)\bigg)
  \nonumber \\
 && = \sum_{\sigma=\pm} \int dx P_t(x,\sigma)
 \bigg(       \sigma v_{\sigma}( x )    \partial_xO(x,\sigma)
    -   \gamma_{\sigma}( x )  O(x,\sigma)
     +   \gamma_{\sigma}( x ) O(x,-\sigma)
     \bigg)
     \nonumber \\
     && \equiv   {\mathbb E}_t  \bigg(       \sigma v_{\sigma}( x )    \partial_xO(x,\sigma)
    -   \gamma_{\sigma}( x )  O(x,\sigma)
     +   \gamma_{\sigma}( x ) O(x,-\sigma)
     \bigg)
\label{runobsdyn}
\end{eqnarray}
The rewriting in the matrix form
\begin{eqnarray}
 \partial_t
{\mathbb E}_t  \begin{pmatrix} 
 O(x,+)
 \\ O(x,-)
  \end{pmatrix}
= {\mathbb E}_t \left[
 \begin{pmatrix} 
-   \gamma_+( x ) + v_+( x )  \partial_x            &     \gamma_+( x )   \\
\gamma_-( x )  & -   \gamma_-( x )  -  v_-( x )  \partial_x          
 \end{pmatrix} 
  \begin{pmatrix} 
 O(x,+)
 \\  O(x,-) 
  \end{pmatrix}
  \right]
  \equiv
{\mathbb E}_t  \left[      {\cal L}^{\dagger}
  \begin{pmatrix} 
 O(x,+)
 \\  O(x,-) 
  \end{pmatrix}
    \right]
\label{runobsadj}
\end{eqnarray}
involves the adjoint operator
\begin{eqnarray}
  {\cal L}^{\dagger}
  = \begin{pmatrix} 
-   \gamma_+( x ) + v_+( x )  \partial_x            &     \gamma_+( x )   \\
\gamma_-( x )  & -   \gamma_-( x )  -  v_-( x )  \partial_x          
 \end{pmatrix} 
 \label{runadjoint}
\end{eqnarray}
of the generator ${\cal L} $ of Eq. \ref{rungenerator}.


\subsection{ Explicit form of the scale vector $\begin{pmatrix} 
 s_+( x)
 \\ s_-( x)
  \end{pmatrix}  $ annihilated by the adjoint ${\cal L}^{\dagger} $
  and independent from $\begin{pmatrix} 
1
 \\ 1
  \end{pmatrix}$ }

 The conservation of the total probability
 \begin{eqnarray}
 1 = \sum_{\sigma=\pm} \int dx  P_t(x,\sigma) =  {\mathbb E}_t [ 1  ]
\label{runnorma}
\end{eqnarray}
 means that the averaged value of observable $O(x,\sigma)=1$
 is conserved by the dynamics and that  $O(x,\sigma)=1$
 is annihilated by the adjoint operator ${\cal L}^{\dagger} $ of Eq. \ref{runadjoint}
    \begin{eqnarray}
0 =  {\cal L}^{\dagger}\begin{pmatrix} 
1
 \\ 1
  \end{pmatrix}  
 \label{runadjoint1}
\end{eqnarray}
  
 Besides this trivial solution, it is interesting to find the other non-trivial solutions 
  $\begin{pmatrix} 
 s_+( x)
 \\ s_-( x)
  \end{pmatrix}  $
  that are annihilated by the adjoint operator $ {\cal L}^{\dagger}$
  since their averaged values will be also conserved by the dynamics
  \begin{eqnarray}
0 =  {\cal L}^{\dagger}\begin{pmatrix} 
 s_+( x)
 \\ s_-( x)
  \end{pmatrix}  
  = \begin{pmatrix} 
-   \gamma_+( x ) + v_+( x )  \partial_x            &     \gamma_+( x )   \\
\gamma_-( x )  & -   \gamma_-( x )  -  v_-( x )  \partial_x          
 \end{pmatrix} \begin{pmatrix} 
 s_+( x)
 \\ s_-( x)
  \end{pmatrix}  
 \label{runadjoints}
\end{eqnarray}
leading to the first-order differential equations
 \begin{eqnarray}
 s_+'( x)   &&=  \frac{   \gamma_+( x ) }{v_+( x )}  \left[ s_+( x) - s_-( x)  \right]         
\nonumber  \\
  s_-'( x)   &&=  \frac{   \gamma_-( x ) }{v_-( x )}  \left[ s_+( x) - s_-( x)  \right]   
 \label{derispm}
\end{eqnarray}

As a consequence, their difference  
\begin{eqnarray}
s(x)  \equiv  s_+( x) - s_-( x)         
 \label{defsdiff}
\end{eqnarray}
satisfies the closed differential equation
\begin{eqnarray}
 s'(x) && =
 s_+'( x)  - s_-'( x)  =    \left( \frac{   \gamma_+( x ) }{v_+( x )} - \frac{\gamma_-( x ) }{v_-( x )}   \right)    s(x)     
 \label{deris}
\end{eqnarray}

When the run-and-tumble dynamics is defined on the whole line $x \in ]-\infty,+\infty[$,
one can choose for instance the particular solution satisfying $s(x=0)=1$ and $s_+(x=0)=0$
\begin{eqnarray}
 s(x) && =  e^{\displaystyle  \int_0^x dy \left[ \frac{ \gamma_+(y)    }{v_+(y) } - \frac{ \gamma_-(y) }{v_-( y)}  \right] }    
 \nonumber \\
 s_+(x) &&= \int_0^x dz \frac{   \gamma_+( z ) }{v_+( z )} s(z)
 =  \int_0^x dz \frac{   \gamma_+( z ) }{v_+( z )}  e^{\displaystyle  \int_0^z dy \left[ \frac{ \gamma_+(y)    }{v_+(y) } - \frac{ \gamma_-(y) }{v_-( y)}  \right] } 
 \label{sx}
\end{eqnarray}
while the corresponding $s_-(x)= s_+(x)-s(x) $ can be then rewritten in various forms
\begin{eqnarray}
 s_-(x) && = s_+(x)-s(x) 
 =   \int_0^x dz \left( \frac{   \gamma_-( z ) }{v_-( z )}+ \frac{   \gamma_+( z ) }{v_+( z )} - \frac{   \gamma_-( z ) }{v_-( z )} \right) e^{\displaystyle  \int_0^z dy \left[ \frac{ \gamma_+(y)    }{v_+(y) } - \frac{ \gamma_-(y) }{v_-( y)}  \right] } -s(x) 
 \nonumber \\
 && = \int_0^x dz  \frac{   \gamma_-( z ) }{v_-( z )} s(z) + \left[ s(z) \right]_0^x -s(x)
 = \int_0^x dz  \frac{   \gamma_-( z ) }{v_-( z )} s(z)  -s(0)
 = \int_0^x dz  \frac{   \gamma_-( z ) }{v_-( z )} s(z)  -1
 \label{sxm}
\end{eqnarray}
When the run-and-tumble dynamics is defined on the half-line $x \in ]0,+\infty[$
or other domains, one can make other choices that are more appropriate for the specific examples that will be described in section \ref{sec_UniversalityClasses}.

In summary,
this non-trivial solution $\begin{pmatrix} 
 s_+( x)
 \\ s_-( x)
  \end{pmatrix}  $ annihilated by the adjoint operator ${\cal L}^{\dagger} $ 
  can be considered as a scale vector
  that generalizes the single scale function $s(x)$ for diffusion processes in the mathematical literature \cite{BookKarlin,BookYor,HandBook,BookBorodin},
  and that governs the exit probabilities from an interval as described in the next subsection.


\subsection{ Exit probabilities $E^{\pm}_{[a,b]}(x,\sigma)$ from an interval $[a,b]$ 
when starting in the state $(x,\sigma)$}

When starting in state $(x,\sigma)$ with $x \in [a,b]$,
the probability $E^+_{[a,b]}(x,\sigma)$ that the first exit from the interval $[a,b]$ 
is via the upper boundary $b$ 
is annihilated by the adjoint ${\cal L}^{\dagger} $
(see the textbooks \cite{gardiner,vankampen,risken})
and can be thus written as a linear combination of the trivial solution $\begin{pmatrix} 
 1
 \\ 1
  \end{pmatrix}  $ 
and of the non-trivial solution $\begin{pmatrix} 
 s_+( x)
 \\ s_-( x)
  \end{pmatrix}  $
discussed in the previous subsection
  \begin{eqnarray}
\begin{pmatrix} 
 E^+_{[a,b]}(x,\sigma=+)
 \\ E^+_{[a,b]}(x,\sigma=-)
  \end{pmatrix} = c_1 \begin{pmatrix} 
1
 \\ 1
  \end{pmatrix}+ c_2 \begin{pmatrix} 
 s_+( x)
 \\ s_-( x)
  \end{pmatrix}  
 \label{ExitCL}
\end{eqnarray}
where the two constants $(c_1,c_2)$ have to be determined by 
the two boundary conditions
\begin{eqnarray}
1 && = E^{+}_{[a,b]}(x=b,\sigma=+) = c_1+ c_2 s_+(b)
\nonumber \\
0 && = E^{+}_{[a,b]}(x=a,\sigma=-)  =c_1+ c_2 s_-(a)
\label{BCExit}
\end{eqnarray}
Note that these boundary conditions involve only the components that 
 have the appropriate internal state to cross the boundary,
 namely $\sigma=+$ at $x=b$ to exit via $b$, and $\sigma=-$ at $x=a$ to exit via $a$,
while the components 
$E^{+}_{[a,b]}(x=b,\sigma=-) $ and $E^{+}_{[a,b]}(x=a,\sigma=+) $ 
that have the wrong internal state to exit, 
are not given a priori but have to be determined by the dynamics.

The solution 
\begin{eqnarray}
E^{+}_{[a,b]}(x,\sigma=+) && = \frac{s_+(x)-s_-(a)}{s_+(b)-s_-(a)}
\nonumber \\
E^{+}_{[a,b]}(x,\sigma=-) && =  \frac{s_-(x)-s_-(a)}{s_+(b)-s_-(a)}
\label{Exit+}
\end{eqnarray}
together with the complementary probabilities $E^-_{[a,b]}(x,\sigma)= 1- E^+_{[a,b]}(x,\sigma)$ to first exit the interval $[a,b]$ via $a$  when starting in state $(x,\sigma)$
\begin{eqnarray}
E^{-}_{[a,b]}(x,\sigma=+) = 1- E^{+}_{[a,b]}(x,\sigma=+) && = \frac{s_+(b)-s_+(x)}{s_+(b)-s_-(a)}
\nonumber \\
E^{-}_{[a,b]}(x,\sigma=-) = 1- E^{+}_{[a,b]}(x,\sigma=-) && =   \frac{s_+(b)-s_-(x)}{s_+(b)-s_-(a)}
\label{Exit-}
\end{eqnarray}
can be summarized by the formula
\begin{eqnarray}
E^{+}_{[a,b]}(x,\sigma) && = \frac{s_{\sigma}(x)-s_-(a)}{s_+(b)-s_-(a)}
\nonumber \\
E^{-}_{[a,b]}(x,\sigma) && =  \frac{s_+(b)-s_{\sigma}(x)}{s_+(b)-s_-(a)}
\label{Exit+-}
\end{eqnarray}
that are the counterpart of the standard result for diffusion processes
without internal state $\sigma$
that involves the single scale function $s(x)$ \cite{c_RGdiff}
\begin{eqnarray}
\text{ Comparison with diffusion processes }: \ \ 
E^{+}_{[a,b]}(x) && = \frac{s(x)-s(a)}{s(b)-s(a)}
\nonumber \\
E^{-}_{[a,b]}(x) && =  \frac{s(b)-s(x)}{s(b)-s(a)}
\label{Exit+-Diff}
\end{eqnarray}


\subsection{ The possible steady states $\begin{pmatrix} 
 P_*(x,+)
 \\ P_*(x,-)
  \end{pmatrix}  $ annihilated by the generator ${\cal L} $ }
  
  The possible steady states of Eq. \ref{run} satisfy
  \begin{eqnarray}
 0=\partial_t P_*(x,+)   &&   =  -    \partial_x  \left[ v_+( x ) P_*(x,+)  \right]    -   \gamma_+( x ) P_*(x,+)   +   \gamma_-( x ) P_*(x,-) 
\nonumber \\
0= \partial_t P_*(x,-)  &&   =   -  \partial_x   \left[  -  v_-( x ) P_*(x,-)  \right]    +   \gamma_+( x ) P_*(x,+)   -   \gamma_-( x ) P_*(x,-) 
\label{runsteady}
\end{eqnarray}
while the steady continuity equation of Eq. \ref{runcontinuity}
\begin{eqnarray}
 0= -    \partial_x J_*(x)
\label{runcontinuitystaedy}
\end{eqnarray}
yields that the steady current $J_*(x) $ of Eq. \ref{runcurrent}
cannot depend on $x$
\begin{eqnarray}
J_* \equiv v_+( x ) P_*(x,+)  -  v_-( x ) P_*(x,-) 
\label{steadycurrent}
\end{eqnarray}

One obtains that the family of possible steady states 
 can be written in terms of the scale functions $s_{\pm}(x)$ and $s(x)$ 
of Eqs \ref{sx}
and \ref{sxm}
as
   \begin{eqnarray}
 P_*(x,+)   &&   =  \frac{ C-J_* s_-(x) }{ v_+(x) s(x) } \geq 0
\nonumber \\
P_*(x,-)   &&   =  \frac{ C-J_* s_+(x) }{ v_-(x) s(x) } \geq 0
\label{Psteady}
\end{eqnarray}
where the corresponding steady current $J_*$ 
and the other constant $C$ associated to the zero-current-solution
will be determined by boundary conditions
and positivity  constraints in specific examples, and one can then discuss whether there exists a normalizable steady state or not.


\section{ Bond probabilities of the renormalized trajectories at scale $\Gamma$ }

\label{sec_bond}

\subsection{ Direct computation of the bond probabilities via probability analysis }

\subsubsection{ Probability $P_{\Gamma}^+(x \vert x_0) $ that an ascending bond 
at scale $\Gamma$ starting at $x_0$ ends at the position $x$ with $x \in [x_0+\Gamma,+\infty[$}

After the minimum at $x_0$,
the probability to have an ascending bond ending at a position bigger than $x$ without any drawdown bigger than $\Gamma$ has the survival form
\begin{eqnarray}
C_{\Gamma}^+(x \vert x_0) \equiv \int_x^{+\infty} dy P_{\Gamma}^+(y \vert x_0) =
 e^{ - \int_{x_0+\Gamma}^x dz R_{\Gamma}^+(z)  } \ \ \ \ \text{ for } \ \ x \in [x_0+\Gamma,+\infty[
  \label{Cplus}
\end{eqnarray}
where $dz R_{\Gamma}^+(z) $ represents the probability to kill the running maximum in state $(z,+)$ by a drawdown $\Gamma$,
i.e. it is given by the exit probability of Eq. \ref{Exit-} 
 from the interval $[a=z-\Gamma,b=z]$ when starting at $(z-dz,\sigma=+)$
\begin{eqnarray}
E^{-}_{[z-\Gamma,z]}(z-dz,\sigma=+) =  \frac{s_+(z)-s_+(z-dz)}{s_+(z)-s_-(z-\Gamma)}
 = \frac{ dz s_+'(z)}{s_+(z)-s_-(z-\Gamma)} \equiv dz R_{\Gamma}^+(z)
\label{Exit-asc}
\end{eqnarray}
with
\begin{eqnarray}
R_{\Gamma}^+(z) \equiv \frac{  s_+'(z)}{s_+(z)-s_-(z-\Gamma)} 
\label{Rxplus}
\end{eqnarray}

The probability distribution $P_{\Gamma}^+(x \vert x_0) $ itself is obtained from
the cumulative distribution of Eq. \ref{Cplus}
via derivation 
\begin{eqnarray}
P_{\Gamma}^+(x \vert x_0) = - \partial_x C_{\Gamma}^+(x \vert x_0) 
= R_{\Gamma}^+(x) e^{ - \int_{x_0+\Gamma}^x dz R_{\Gamma}^+(z)  } \ \ \ \ \text{ for } \ \ x \in [x_0+\Gamma,+\infty[
  \label{Pplus}
\end{eqnarray}

In particular, the probability of an ascending bond that is about to be decimated at scale $\Gamma$ reduces to
\begin{eqnarray}
P_{\Gamma}^+(x=x_0+\Gamma \vert x_0) 
= R_{\Gamma}^+(x_0+\Gamma) = \frac{ s_+'(x_0+\Gamma) }{ s_+(x_0+\Gamma)-s_-(x_0) } 
= \partial_{\Gamma} \ln \left[ s_+(x_0+\Gamma)-s_-(x_0) \right]
  \label{Pplusdecim}
\end{eqnarray}


\subsubsection{ Probability $P_{\Gamma}^-(x \vert x_0) $ that a descending bond 
at scale $\Gamma$ starting at $x_0$ ends at the position $x$ with $x \in ]-\infty,x_0-\Gamma]$}

Similarly after a maximum at $x_0$, 
the probability to have a descending bond ending at a position lower than $x$ without any drawup bigger than $\Gamma$ has the survival form
\begin{eqnarray}
C_{\Gamma}^-(x \vert x_0) \equiv \int_{-\infty}^x dy P_{\Gamma}^-(y \vert x_0)
 =
 e^{ - \int_x^{x_0-\Gamma} dz R_{\Gamma}^-(z)  } \ \ \ \ \text{ for } \ \ x \in ]-\infty,x_0-\Gamma]
  \label{Cmoins}
\end{eqnarray}
where $dz R_{\Gamma}^-(z) $ represents the probability to kill the running minimum in state $(z,-)$ by a drawup $\Gamma$,
i.e. it is given by the exit probability of Eq. \ref{Exit+} 
 from the interval $[a=z,b=z+\Gamma]$ when starting at $(z+dz,\sigma=-)$
\begin{eqnarray}
E^{+}_{[z,z+\Gamma]}(z+dz,\sigma=-) && =  \frac{s_-(z+dz)-s_-(z)}{s_+(z+\Gamma)-s_-(z)}
=  \frac{ dz s_-'(z)}{s_+(z+\Gamma)-s_-(z)}\equiv dz R_{\Gamma}^-(z)
\label{Exit+des}
\end{eqnarray}
with
\begin{eqnarray}
R_{\Gamma}^-(z) \equiv \frac{  s_-'(z)}{s_+(z+\Gamma)-s_-(z)}
\label{Rxmoins}
\end{eqnarray}

The probability distribution $P_{\Gamma}^-(x \vert x_0) $ itself is obtained 
from the cumulative distribution of Eq. \ref{Cmoins}
via derivation 
\begin{eqnarray}
P_{\Gamma}^-(x \vert x_0) =  \partial_x C_{\Gamma}^-(x\vert x_0) 
= R_{\Gamma}^-(x) e^{ - \int_x^{x_0-\Gamma} dz R_{\Gamma}^-(z)  } \ \ \ \ \text{ for } \ \  x \in ]-\infty,x_0-\Gamma]
  \label{Pmoins}
\end{eqnarray}

In particular, the probability of a descending bond that is about to be decimated at scale $\Gamma$ reduces to
\begin{eqnarray}
P_{\Gamma}^-(x=x_0-\Gamma \vert x_0) 
= R_{\Gamma}^-(x_0-\Gamma) =  \frac{ s_-'(x_0-\Gamma) }{ s_+(x_0)-s_-(x_0-\Gamma) }
=  \partial_{\Gamma} \ln \left[ s_+(x_0)-s_-(x_0-\Gamma) \right]
  \label{Pmoinsdecim}
\end{eqnarray}


\subsection{ Analysis via the RG-flows as a function of the renormalization scale $\Gamma$ }

\subsubsection{RG-flows for the functions $R^{\pm}_{\Gamma}(x) $   }

In the previous work \cite{c_RGdiff} concerning diffusion processes, it was shown that the RG-flows for the bond probabilities $P^{\pm}_{\Gamma}(x \vert x_0)  $ given by the Ansatz of Eqs \ref{Pplus} and \ref{Pmoins}
are satisfied if the two functions $R^{\pm}_{\Gamma}(x) $ 
evolve according to
\begin{eqnarray}
  \partial_{\Gamma}   \ln R_{\Gamma}^+(x)    && =-  R_{\Gamma}^-( x-\Gamma) 
    \nonumber \\
\partial_{\Gamma}  \ln R_{\Gamma}^-(x) && = - R_{\Gamma}^+( x+\Gamma)
  \label{RGforRpm}
\end{eqnarray}

These RG-flows of \ref{RGforRpm}
are indeed satisfied by the explicit forms of Eqs \ref{Rxplus}
and \ref{Rxmoins} 
\begin{eqnarray}
 \partial_{\Gamma}   \ln R^+_{\Gamma}(x) = \partial_{\Gamma}   \ln \frac{ s_+'(x) }{ s_+(x)-s_-(x-\Gamma) } 
= - \frac{ s_-'(x-\Gamma) }{ s_+(x)-s_-(x-\Gamma) }  =-  R_{\Gamma}^-( x-\Gamma) 
  \label{Rxplusln}
\end{eqnarray}
and
\begin{eqnarray}
 \partial_{\Gamma}   \ln R^-_{\Gamma}(x) =  \partial_{\Gamma}   \ln  \frac{ s_-'(x) }{ s_+(x+\Gamma)-s_-(x) } 
= -  \frac{ s_+'(x+\Gamma) }{ s_+(x+\Gamma)-s_-(x) } =- R_{\Gamma}^+( x+\Gamma)
  \label{Rxmoinsln}
\end{eqnarray}


\subsubsection{Exact initial condition at $\Gamma=0$ corresponding to the decomposition of the initial trajectory into its ballistic intervals between switching events   }

An important novelty with the case of the renormalization of diffusion 
trajectories \cite{c_RGdiff} is that the bond distributions $P_{\Gamma}^{\pm}(x \vert x_0) $ of Eqs \ref{Pplus}
and \ref{Pmoins} can be written even for 
the vanishing RG-scale $\Gamma=0$ corresponding to the initial 
run-and-tumble trajectory
that is already an alternating sequence of ascending and descending bonds
 \begin{eqnarray}
P_{\Gamma=0}^+(x \vert x_0) 
&&= R_0^+(x) e^{ - \int_{x_0}^x dz R_0^+(z)  } \ \ \ \ \text{ for } \ \ x \in [x_0,+\infty[
 \nonumber \\
 P_{\Gamma=0}^-(x \vert x_0) 
&& = R_0^-(x) e^{ - \int_x^{x_0} dz R_0^-(z)  } \ \ \ \ \text{ for } \ \  x \in ]-\infty,x_0]
  \label{PGamma0}
\end{eqnarray}
where the two functions $R_{\Gamma=0}^{\pm}(z) $ of Eqs \ref{Rxplus}
and \ref{Rxmoins}
simplify for $\Gamma=0$ using Eq. \ref{derispm} 
\begin{eqnarray}
R^+_{\Gamma=0}(z) &&= \frac{  s_+'(z)}{s_+(z)-s_-(z)} =\frac{s_+'(z)}{s(z)} = \frac{\gamma_+( z )}{v_+(z)}
\nonumber \\
R_{\Gamma=0}^-(z) && 
=  \frac{  s_-'(z)}{s_+(z)-s_-(z)} = \frac{  s_-'(z)}{s(z)}  =\frac{\gamma_-( z )}{v_-(z)}
\label{RGamma0}
\end{eqnarray}

Plugging these values into Eq. \ref{PGamma0} yields
\begin{eqnarray}
P^+_{\Gamma=0}(x \vert x_0) 
= \frac{\gamma_+(x) }{v_+(x)} \ e^{\displaystyle - \int_{x_0}^{x} dz \frac{\gamma_+( z )}{v_+(z)}  }
\ \ \ \ \text{ for } \ \ x \in [x_0,+\infty[
\nonumber \\
P^-_{\Gamma=0}(x \vert x_0) 
= \frac{\gamma_-(x) }{v_-(x)} \ e^{\displaystyle - \int_{x}^{x_0} dz \frac{\gamma_-( z )}{v_-(z)}  }
\ \ \ \ \text{ for } \ \  x \in ]-\infty,x_0]
\label{PIniBallisticIntervals}
\end{eqnarray}
that characterize the ballistic intervals between switching events 
for the initial run-and-tumble trajectories \cite{c_largeDevRun}.

The fact that this exact initial condition for the bond probabilities $P^{\pm}_{\Gamma=0}(x \vert x_0)  $ at $\Gamma=0$
depend only on the two ratios $\frac{\gamma_+( z ) }{v_+( z )}$ and $\frac{\gamma_-( z ) }{v_-( z )} $ shows that these two ratios determine the universality class
of the renormalized trajectories
when one considers only the amplitudes as in all sections of the 
main text, while the additional duration variables 
that will be included in the Appendices involve the velocities $v_{\pm}$
themselves (see the discussion around Eqs \ref{durationsinter} concerning the durations associated to the initial condition of Eq. \ref{PIniBallisticIntervals}).


\section{ Densities $n_{\Gamma}^{\pm}$ of minima and maxima in the renormalized trajectories }

\label{sec_nextrema}

\subsection{  Equations satisfied by the densities $n_{\Gamma}^{\pm}$ of 
maxima and minima  }

\subsubsection{ Densities $n_{\Gamma}^{\pm}(x)$ of  extrema
should be stable by the Markov chain generated by the conditional probabilities $P_{\Gamma}^{\pm}(x \vert x_0) $ }

As discussed in detail in \cite{c_RGdiff}, once the bond probabilities $P_{\Gamma}^{\pm}(x \vert x_0) $ have been obtained, the density $n_{\Gamma}^{+}(x)$ of maxima and the density $n_{\Gamma}^{-}(x)$ of minima at scale $\Gamma$
can be obtained via their 
stationarity under the Markov chain generated by the conditional probabilities $P_{\Gamma}^{\pm}(x \vert x_0) $ of Eqs \ref{Pplus}
and \ref{Pmoins}
that play the role of transition matrices
\begin{eqnarray}
n_{\Gamma}^{+}(x) && = \int_{-\infty}^{x-\Gamma} dx_0 P_{\Gamma}^+(x \vert x_0) n_{\Gamma}^-(x_0)  
= R_{\Gamma}^+(x)  \int_{-\infty}^{x-\Gamma} dx_0 e^{ - \int_{x_0+\Gamma}^x dz R_{\Gamma}^+(z)  } n_{\Gamma}^-(x_0)  
\nonumber \\
n_{\Gamma}^{-}(x)&& = \int_{x+\Gamma}^{+\infty} dx_0 P_{\Gamma}^-(x \vert x_0) n_{\Gamma}^+(x_0)
= R_{\Gamma}^-(x) \int_{x+\Gamma}^{+\infty} dx_0 e^{ - \int_x^{x_0-\Gamma} dz R_{\Gamma}^-(z)  }n_{\Gamma}^+(x_0)
  \label{MarkovChain}
\end{eqnarray}


\subsubsection{ RG-flows satisfied by the densities $n_{\Gamma}^{\pm}(x)$ of extrema }

As discussed in detail in \cite{c_RGdiff},
the RG-flows satisfied by the densities $n_{\Gamma}^{\pm}(x)$ of
minima and maxima read
\begin{eqnarray}
\partial_{\Gamma} n_{\Gamma}^-(x) &&   =  - n_{\Gamma}^-(x)  
P_{\Gamma}^+( x+\Gamma \vert x )
 - n_{\Gamma}^+(x+\Gamma) 
 P_{\Gamma}^-( x \vert x+\Gamma)
 \nonumber \\
 \partial_{\Gamma} n_{\Gamma}^+(x) &&    
 =  - n_{\Gamma}^+(x) 
 P_{\Gamma}^-( x-\Gamma \vert x ) 
 - n_{\Gamma}^-(x-\Gamma) 
 P_{\Gamma}^+( x \vert x-\Gamma)
  \label{RGnmaxforward}
\end{eqnarray}
where one can plug the probabilities of about-to-be-decimated bonds of Eqs \ref{Pplusdecim}
and \ref{Pmoinsdecim} to obtain their explicit forms in terms
of the pair of scale functions $s_{\pm}(x)$
\begin{eqnarray}
\partial_{\Gamma} n_{\Gamma}^-(x) &&   
=  - n_{\Gamma}^-(x)   
\frac{ s_+'(x+\Gamma) }{ s_+(x+\Gamma)-s_-(x) }
 - n_{\Gamma}^+(x+\Gamma)  
 \frac{ s_-'(x) }{ s_+(x+\Gamma)-s_-(x) }
 \nonumber \\
 \partial_{\Gamma} n_{\Gamma}^+(x) &&   
  =  - n_{\Gamma}^+(x)  
  \frac{ s_-'(x-\Gamma) }{ s_+(x)-s_-(x-\Gamma) }
 - n_{\Gamma}^-(x-\Gamma)
 \frac{ s_+'(x) }{ s_+(x)-s_-(x-\Gamma) }
  \label{RGnmaxexpli}
\end{eqnarray}


\subsection{  General solutions involving two constants $(c,\kappa)$  }

Using the same methods as in \cite{c_RGdiff}, 
one obtains that the general solutions of Eqs \ref{MarkovChain}
and \ref{RGnmaxexpli}
\begin{eqnarray}
n^+_{\Gamma}(x) && =\frac{s_+'(x) \left[c-\kappa s_-(x-\Gamma)\right]}{\left[s_+(x)-s_-(x-\Gamma)\right]^2} 
\nonumber \\
n^-_{\Gamma}(x) && =\frac{s_-'(x) \left[c-\kappa s_+(x+\Gamma)\right]}{\left[s_+(x+\Gamma)-s_-(x)\right]^2} 
\label{densities}
\end{eqnarray}
depend on two constants $(c,\kappa)$,
that can be determined in the present case
by the explicit initial condition at $\Gamma=0$
 using Eqs  \ref{derispm}
\begin{eqnarray}
n^+_{\Gamma=0}(x) && =\frac{s_+'(x) \left[c-\kappa s_-(x)\right]}{\left[s_+(x)-s_-(x)\right]^2} 
= \frac{ \frac{   \gamma_+( x ) }{v_+( x )} \left[c-\kappa s_-(x)\right]}{s(x)} 
\nonumber \\
n^-_{\Gamma=0}(x) && =\frac{s_-'(x) \left[c-\kappa s_+(x)\right]}{\left[s_+(x)-s_-(x)\right]^2} 
=\frac{\frac{   \gamma_-( x ) }{v_-( x )} \left[c-\kappa s_+(x)\right]}{s(x)} 
\label{densities0}
\end{eqnarray}
that should correspond to the densities of switching events in the steady state of Eq. \ref{Psteady}
   \begin{eqnarray}
\gamma_+( x ) P_*(x,+)   &&   = \gamma_+( x ) \frac{ C-J_* s_-(x) }{ v_+(x) s(x) } 
\nonumber \\
\gamma_-( x )P_*(x,-)   &&   = \gamma_-( x ) \frac{ C-J_* s_+(x) }{ v_-(x) s(x) } 
\label{FlowsPsteady}
\end{eqnarray}
leading to the identification of the two constants $c=C$ and $\kappa=J_*$.
\begin{eqnarray}
c && =C
\nonumber \\
\kappa && =J_*
\label{Qsteady}
\end{eqnarray}
  Since the parameter $\kappa=J_*$ has the physical meaning of a steady current,
  there are special properties when this parameter vanishes as discussed in the next subsection.


\subsection{ Case $\kappa=J_*=0$ : Detailed-Balance for the renormalized trajectories   }

In this subsection, we focus on the vanishing-current case $\kappa=J=0$,
where the densities of extrema of Eq. \ref{densities}
reduce to
\begin{eqnarray}
n^+_{\Gamma}(x) && =c \frac{s_+'(x) }{\left[s_+(x)-s_-(x-\Gamma)\right]^2} 
\nonumber \\
n^-_{\Gamma}(x) && =c \frac{s_-'(x) }{\left[s_+(x+\Gamma)-s_-(x)\right]^2} 
\label{densitiesJ0}
\end{eqnarray}


\subsubsection{ Detailed-Balance properties of the conditional probabilities $P_{\Gamma}^{\pm}(. \vert .) $ 
with respect to the densities $n_{\Gamma}^{\pm}$ of extrema }

The two functions $R^{\pm}_{\Gamma}(x)$ introduced in Eqs \ref{Rxplus}
and \ref{Rxmoins}
respectively are related by
\begin{eqnarray}
\frac{ R^-_{\Gamma}(x) }{ s_-'(x) } =  \frac{ 1 }{ s_+(x+\Gamma)-s_-(x) } = \frac{ R^+_{\Gamma}(x+\Gamma) }{ s_+'(x+\Gamma) } 
  \label{Rpm}
\end{eqnarray}

As a consequence, the integral of $R_{\Gamma}^-(z) $ that appears in the exponential of the conditional probability 
$P_{\Gamma}^-(x \vert x_0)$ of Eq. \ref{Pmoins} for descending bonds
\begin{eqnarray}
 \int_x^{x_0-\Gamma} dz R_{\Gamma}^-(z)
&& = \int_x^{x_0-\Gamma} dz \frac{ s_-'(z) }{ s_+(z+\Gamma)-s_-(z) }   
  =  \int_{x+\Gamma}^{x_0} dy \frac{ s_-'(y-\Gamma) }{ s_+(y)-s_-(y-\Gamma) }   
  \nonumber \\
&& =  
\int_{x+\Gamma}^{x_0} dy \left( \frac{ s_+'(y)}{ s_+(y)-s_-(y-\Gamma) } 
- \frac{ s_+'(y) - s_-'(y-\Gamma) }{ s_+(y)-s_-(y-\Gamma) }  \right) 
= 
 \int_{x+\Gamma}^{x_0} dy R_{\Gamma}^+(y)  
  -  \left[ \ln \left( s_+(y)-s_-(y-\Gamma) \right) \right]_{x+\Gamma}^{x_0}
   \nonumber \\
&& = \int_{x+\Gamma}^{x_0} dy R_{\Gamma}^+(y)  
 +  \ln\left( \frac{  s_+(x+\Gamma)-s_-(x)  }{  s_+(x_0)-s_-(x_0-\Gamma)  } \right)
  \label{DBintegR}
\end{eqnarray}
   can be thus rewritten, up to boundary terms,
as an integral of the function $R_{\Gamma}^+(y) $ that appears in the 
exponential of the conditional probability 
$P_{\Gamma}^+(x_0 \vert x)$ of Eq. \ref{Pplus} for ascending bonds where the two end points $(x_0,x)$ are exchanged
\begin{eqnarray}
P_{\Gamma}^+(x_0 \vert x)  
= R_{\Gamma}^+(x_0) e^{ - \int_{x+\Gamma}^{x_0} dy R_{\Gamma}^+(y)  } 
= \frac{ s_+'(x_0) }{ s_+(x_0)-s_-(x_0-\Gamma) }  e^{ - \int_{x+\Gamma}^{x_0} dy R_{\Gamma}^+(y)  }
  \label{Pplusx0xexchanged}
\end{eqnarray}

Plugging Eq. \ref{DBintegR} into Eq. \ref{Pmoins} yields
that $P_{\Gamma}^-(x \vert x_0) $ 
can be rewritten in terms of $P_{\Gamma}^+(x_0 \vert x) $
 \begin{eqnarray}
P_{\Gamma}^-(x \vert x_0) 
&& = R_{\Gamma}^-(x) e^{ - \int_x^{x_0-\Gamma} dz R_{\Gamma}^-(z)  } 
=\frac{ s_-'(x) }{ s_+(x+\Gamma)-s_-(x) } 
e^{-  \int_x^{x_0-\Gamma} dz R_{\Gamma}^-(z) }
\nonumber \\
&&=\frac{ s_-'(x) }{ s_+(x+\Gamma)-s_-(x) } e^{- \int_{x+\Gamma}^{x_0} dy R_{\Gamma}^+(y)   -  \ln\left( \frac{  s_+(x+\Gamma)-s_-(x)  }{  s_+(x_0)-s_-(x_0-\Gamma)  } \right)}
=\frac{ s_-'(x) [ s_+(x_0)-s_-(x_0-\Gamma)] }{ [s_+(x+\Gamma)-s_-(x)]^2 } e^{- \int_{x+\Gamma}^{x_0} dy R_{\Gamma}^+(y)  }
\nonumber \\
&&=\frac{ s_-'(x) [ s_+(x_0)-s_-(x_0-\Gamma)]^2 }{ s_+'(x_0) [s_+(x+\Gamma)-s_-(x)]^2 } P_{\Gamma}^+(x_0 \vert x) 
\equiv \frac{ n_{\Gamma}^-(x) }{ n_{\Gamma}^+(x_0)} P_{\Gamma}^+(x_0 \vert x)
  \label{PmoinsDB}
\end{eqnarray}
where one recognizes the density $n_{\Gamma}^+(x_0) $ of maxima and the density $n_{\Gamma}^-(x) $ of minima  of Eq. \ref{densitiesJ0}. 
The physical meaning of this identity
 \begin{eqnarray}
P_{\Gamma}^-(x \vert x_0) n_{\Gamma}^+(x_0)
= P_{\Gamma}^+(x_0 \vert x)n_{\Gamma}^-(x)
  \label{DBRGwithn}
\end{eqnarray}
is thus the Detailed-Balance of the conditional probabilities $P_{\Gamma}^-(. \vert .) $ at scale $\Gamma$
with respect to the steady densities $n_{\Gamma}^{\pm}$ of extrema of the Markov chain of Eq. \ref{MarkovChain}.


\subsubsection{ Similarity transformation of the conditional probabilities $P_{\Gamma}^{\pm}(. \vert .) $
towards symmetric blocks $ B^{\pm}_{\Gamma}(.,.) $}

Whenever Markov generators satisfy detailed-balance,
they can be transformed into symmetric operators  
via similarity transformations (see the textbooks \cite{gardiner,vankampen,risken}) 
and this property plays an important role in
 many specific applications to various models (see for instance \cite{glauber,Felderhof,siggia,referee,kimball,peschel,jpb_antoine,locgolosov,texier,us_eigenvaluemethod,Castelnovo,c_lyapunov,us_gyrator,us_kemeny,c_pearson,c_boundarydriven,c_largedevDBsusy,c_susyFokkerPlanck}).

As in the previous work \cite{c_RGdiff},
it is thus useful to define the
similarity transformations for the conditional probabilities
satisfying the detailed-balance property of Eqs \ref{DBRGwithn} 
 \begin{eqnarray}
P_{\Gamma}^-(x \vert x_0) && = \sqrt{ n_{\Gamma}^-(x)} B_{\Gamma}^-(x,x_0) \frac{1}{\sqrt{n_{\Gamma}^+(x_0) }  }
\nonumber \\
P_{\Gamma}^+(x \vert x_0) && = \sqrt{ n_{\Gamma}^+(x)} B_{\Gamma}^+(x,x_0) \frac{1}{\sqrt{n_{\Gamma}^-(x_0) }  }
  \label{SimilarityB}
\end{eqnarray}
so that the 
detailed-balance property of Eqs \ref{DBRGwithn} 
rewritten for the blocks $ B_{\Gamma}(.,.) $ 
 \begin{eqnarray}
B_{\Gamma}^-(x,x_0) \sqrt{n_{\Gamma}^+(x_0) n_{\Gamma}^-(x)} 
= P_{\Gamma}^-(x \vert x_0) n_{\Gamma}^+(x_0)
= P_{\Gamma}^+(x_0 \vert x)n_{\Gamma}^-(x)  
= \sqrt{ n_{\Gamma}^+(x_0) n_{\Gamma}^-(x)} B_{\Gamma}^+(x_0,x) 
  \label{SymBlockB}
\end{eqnarray}
yield that they are symmetric with respect to the exchange of their two end-points
 \begin{eqnarray}
  B_{\Gamma}^-(x,x_0) = B_{\Gamma}^+(x_0,x) 
  \label{SymBlockBex}
\end{eqnarray}
as can be also checked via
 their explicit forms obtained from Eq. \ref{PmoinsDB} using Eq. \ref{densitiesJ0} \begin{eqnarray}
B_{\Gamma}^-(x,x_0)  && = \frac{1}{\sqrt{n_{\Gamma}^-(x) }  } P_{\Gamma}^-(x \vert x_0) \sqrt{n_{\Gamma}^+(x_0) }
=  \frac{s_+(x+\Gamma)-s_-(x)}{ \sqrt{ s_-'(x) }}  P_{\Gamma}^-(x \vert x_0)  \frac{ \sqrt{s_+'(x_0)} }{s_+(x_0)-s_-(x_0-\Gamma) }
 \nonumber \\
&& 
=    \frac{ \sqrt{ s_-'(x) s_+'(x_0)} }{s_+(x_0)-s_-(x_0-\Gamma) } e^{-  \int_x^{x_0-\Gamma} dz R_{\Gamma}^-(z) }
=\frac{ \sqrt{ s_-'(x) s_+'(x_0)}  }{ s_+(x+\Gamma)-s_-(x) } e^{- \int_{x+\Gamma}^{x_0} dy R_{\Gamma}^+(y)  }
\nonumber \\
&&=\frac{  [ s_+(x_0)-s_-(x_0-\Gamma)] }{ \sqrt{s_+'(x_0)}  } 
P_{\Gamma}^+(x_0 \vert x) 
\frac{ \sqrt{ s_-'(x) }  }{  [s_+(x+\Gamma)-s_-(x)] }
= \frac{1}{\sqrt{n_{\Gamma}^+(x_0) }  }P_{\Gamma}^+(x_0 \vert x) \sqrt{n_{\Gamma}^{-}(x) } = B_{\Gamma}^+(x_0, x)
  \label{ExpliciBlockB}
\end{eqnarray}
with the special case of blocks about to be decimated
 \begin{eqnarray}
B_{\Gamma}^-(x=x_0-\Gamma,x_0)  && =     \frac{ \sqrt{ s_-'(x_0-\Gamma) s_+'(x_0)} }{s_+(x_0)-s_-(x_0-\Gamma) } 
\nonumber \\
B_{\Gamma}^+(x_0=x+\Gamma, x) && =\frac{ \sqrt{ s_-'(x) s_+'(x+\Gamma)}  }{ s_+(x+\Gamma)-s_-(x) } 
  \label{ExpliciBlockBdecim}
\end{eqnarray}


\section{ Universality classes determined by the two ratios $\frac{\gamma_+( z ) }{v_+( z )}$ and $\frac{\gamma_-( z ) }{v_-( z )} $ with simple examples  }

\label{sec_UniversalityClasses}

\subsection{Universality classes determined by the two ratios $\frac{\gamma_+( z ) }{v_+( z )}$ and $\frac{\gamma_-( z ) }{v_-( z )} $ } 

The universality classes are defined by the pair of scale functions $s_{\pm}(x)$
of Eq. \ref{sx} that determines all properties of the renormalized trajectories
when one considers only the amplitudes as in all sections of the 
main text (while the additional duration variables are included only in the Appendices).
So the universality class is determined by the two ratios 
$\frac{\gamma_+( z ) }{v_+( z )}$ and $\frac{\gamma_-( z ) }{v_-( z )} $.
This property is also obvious from the exact initial condition at $\Gamma=0$
of Eq. \ref{PIniBallisticIntervals}
that involves only these two ratios.

As a consequence in each universality class, 
there are various simplifying models, in particular:

(i) the model where the two velocities are equal to unity $v_{\pm}(x)=1$,
where the time duration of each bond is equal to its length. 

(ii) the model where the two switching rates are equal to unity $\gamma_{\pm}(x)=1$,
where the time counts instead the averaged number of tumbles.


\subsection{ Uniform symmetric Run-and-Tumble Particle $v_{\pm}(x)=v$ and $\gamma_{\pm}(x)=\gamma $ }

\label{subsec_RTPsym}

Let us apply the framework of the previous sections to
the simplest case of the uniform symmetric RTP $v_{\pm}(x)=v$ and $\gamma_{\pm}(x)=\gamma $, where Eqs \ref{sx}
and \ref{sxm} 
\begin{eqnarray}
 s(x) && =  1 
 \nonumber \\
 s_+(x) &&=  \frac{\gamma }{v} x 
\nonumber \\
 s_-(x) && = s_+(x)-s(x) 
 =   \frac{\gamma }{v} x  -1
 \label{sxu}
\end{eqnarray}
lead to the exit probabilities of Eq. \ref{Exit+}
\begin{eqnarray}
E^{+}_{[a,b]}(x,\sigma=+) && = \frac{x-a+\frac{v}{\gamma}}{b-a+\frac{v}{\gamma}}
\nonumber \\
E^{+}_{[a,b]}(x,\sigma=-) && =  \frac{x-a}{b-a+\frac{v}{\gamma}}
\label{Exit+u}
\end{eqnarray}

The functions of Eqs \ref{Rxplus} \ref{Rxmoins} coincide 
\begin{eqnarray}
R_{\Gamma}^{\pm}(z) = \frac{  1}{\Gamma + \frac{v}{\gamma}} 
\label{Rxu}
\end{eqnarray}
and do not depend on $z$, so that the bond probabilities
of Eqs \ref{Pplus}
and \ref{Pmoins}
reduce to exponentials 
\begin{eqnarray}
P_{\Gamma}^+(x \vert x_0) 
= \frac{  1}{\Gamma + \frac{v}{\gamma}}  e^{ -  \frac{  x-(x_0+\Gamma)}{\Gamma + \frac{v}{\gamma}}   } \ \ \ \ \text{ for } \ \ x \in [x_0+\Gamma,+\infty[
\nonumber \\
P_{\Gamma}^-(x \vert x_0) 
= \frac{  1}{\Gamma + \frac{v}{\gamma}}  
e^{ -  \frac{  (x_0-\Gamma)-x}{\Gamma + \frac{v}{\gamma}}   } \ \ \ \ \text{ for } \ \  x \in ]-\infty,x_0-\Gamma]
  \label{P+-u}
\end{eqnarray}
while the densities of extrema of Eq \ref{densitiesJ0} read
\begin{eqnarray}
n^{\pm}_{\Gamma}(x) && =c \frac{\frac{\gamma }{v} }
{\left[ 1+ \frac{\gamma }{v} \Gamma\right]^2}  
= c \frac{\frac{v}{\gamma } }
{\left[ \Gamma+ \frac{v}{\gamma } \right]^2}  
\label{densitiesJ0u}
\end{eqnarray}

The interpretation is thus that the persistence length $l \equiv \frac{v}{\gamma}$ of the run-and-tumble particle dominates the ballistic regime $\Gamma \ll l$ 
but disappears in the diffusive regime $\Gamma \gg l$ where one recovers the standard Brownian result.



\subsection{ Uniform asymmetric Run-and-Tumble Particle $v_{\pm}(x)=1$ and 
$\gamma_{\pm}(x)=\gamma_{\pm} $ with $\gamma \equiv \gamma_+-\gamma_-\ne 0$}

The scale functions of Eqs \ref{sx}
and \ref{sxm} are exponential
\begin{eqnarray}
 s(x) && =  e^{\gamma x}    
 \nonumber \\
 s_+(x) &&= \frac{\gamma_+}{\gamma}    e^{\gamma x}
\nonumber \\
 s_-(x) && =\frac{\gamma_-}{\gamma}    e^{\gamma x}
  \label{sxmdrift}
\end{eqnarray}
and lead to functions of Eqs \ref{Rxplus} \ref{Rxmoins} that do not depend on $z$
\begin{eqnarray}
R_{\Gamma}^+(z) && = \frac{  \gamma_+ \gamma}
{\gamma_+- \gamma_- e^{-\gamma \Gamma}} \equiv R_{\Gamma}^+
\nonumber \\
R_{\Gamma}^-(z) && = \frac{  \gamma_- \gamma}
{\gamma_+ e^{\gamma \Gamma}-\gamma_-} \equiv R_{\Gamma}^-
\label{Rx+-drift}
\end{eqnarray}
so that 
the bond probabilities
of Eqs \ref{Pplus}
and \ref{Pmoins} are exponential
\begin{eqnarray}
P_{\Gamma}^+(x \vert x_0)  
= R_{\Gamma}^+ e^{ - [x-(x_0+\Gamma) ]R_{\Gamma}^+  } \ \ \ \ \text{ for } \ \ x \in [x_0+\Gamma,+\infty[
 \nonumber \\
P_{\Gamma}^-(x \vert x_0)  
= R_{\Gamma}^- e^{ - [ (x_0-\Gamma)-x] R_{\Gamma}^-  } \ \ \ \ \text{ for } \ \  x \in ]-\infty,x_0-\Gamma]
  \label{P+-drift}
\end{eqnarray}

The general expressions for the extrema densities of Eq \ref{densities}
\begin{eqnarray}
n^+_{\Gamma}(x) && =\frac{ \gamma  \gamma_+ 
\left[c \gamma e^{- \gamma x} -\kappa \gamma_-    e^{-\gamma \Gamma }\right]}
{\left[\gamma_+- \gamma_- e^{-\gamma \Gamma}\right]^2} 
\nonumber \\
n^-_{\Gamma}(x) && =\frac{ \gamma  \gamma_-
\left[c \gamma e^{- \gamma x} -\kappa \gamma_+    e^{\gamma \Gamma }\right]}
{\left[\gamma_+ e^{\gamma \Gamma}-\gamma_-\right]^2} 
\label{densitiesdrift}
\end{eqnarray}
yield that the expected translation-invariant physical solutions correspond to 
$c=0$  leading to 
\begin{eqnarray}
n^{\pm}_{\Gamma}(x) && =-\kappa \frac{ \gamma  \gamma_+ 
 \gamma_-    }
{\left[\gamma_+e^{ \frac{\gamma}{2} \Gamma}- \gamma_- e^{- \frac{\gamma}{2} \Gamma}\right]^2} 
\label{densitiesdriftcurrent}
\end{eqnarray}
where the non-vanishing current parameter $\kappa \propto (-\gamma) = \gamma_--\gamma_+$
is associated to the flow towards $(\pm \infty)$,
as in the case of the Brownian with drift discussed in \cite{c_RGdiff}.


\subsection{ Geometric RTP on the half-line $x \in ]0,+\infty[$ with 
constant rates $\gamma_{\pm} (x)=\gamma_{\pm}$  
and linear velocities $v_{\pm}( x )=x $  }

For the run-and-tumble process defined on the half-line $x \in ]0,+\infty[ $
with 
constant rates $\gamma_{\pm} (x)=\gamma_{\pm}$  
and linear velocities $v_{\pm}( x )=x $, the process $y(t)=\ln x(t)$ 
is the asymmetric RTP studied in the previous subsection,
where $\gamma \equiv \gamma_+-\gamma_-$ determines the asymptotic behavior :

(i) for $\gamma<0$, the processes $y(t)=\ln x(t)$ and $x(t)$ both flows towards $(+\infty)$

(ii) for $\gamma>0$, the process $y(t)=\ln x(t)$ flows to $(-\infty)$ and the process $x(t)$ flows to $0^+$

(iii) for $\gamma=0$, the process $y(t)=\ln x(t)$ is the symmetric RTP with $v_{\pm}=1$ discussed in subsection \ref{subsec_RTPsym}.

The scale functions satisfying Eqs \ref{derispm} and \ref{deris}
are given by the following power-laws when $\gamma \ne 0$
(the special case $\gamma=0$ will be discussed in subsection \ref{subsec_gamma0})
\begin{eqnarray}
 s(x) && =   x^{\gamma} 
 \nonumber \\
 s_{+}(x) &&=   \frac{\gamma_+}{\gamma} x^{\gamma}
  \nonumber \\
 s_{-}(x) &&=   \frac{\gamma_-}{\gamma} x^{\gamma}
 \label{sfree}
\end{eqnarray}

The scale functions of Eq. \ref{sfree}
yield that the functions of Eqs \ref{Rxplus} \ref{Rxmoins} 
can be rewritten in scaling forms
\begin{eqnarray}
R_{\Gamma}^+(z) && = \frac{  s_+'(z)}{s_+(z)-s_-(z-\Gamma)} 
=  \frac{  \gamma \gamma_+ z^{\gamma-1}}
{\gamma_+ z^{\gamma}-  \gamma_- (z-\Gamma)^{\gamma}
}  \equiv   \frac{ 1 }{ \Gamma  } r^+ \left( \frac{z}{\Gamma} \right)
\ \ \text{ for } \ \ z \in ]\Gamma,+\infty[
\nonumber \\
R_{\Gamma}^-(z) && = \frac{  s_-'(z)}{s_+(z+\Gamma)-s_-(z)}
= \frac{ \gamma \gamma_- z^{\gamma-1}}
{\gamma_+ (z+\Gamma)^{\gamma}-  \gamma_- z^{\gamma}
} \equiv   \frac{ 1 }{ \Gamma  } r^- \left( \frac{z}{\Gamma} \right)
\ \ \text{ for } \ \ z \in ]0,+\infty[
\label{Rx+-free}
\end{eqnarray}

As a consequence, the bond probabilities of Eqs \ref{Pplus} and \ref{Pmoins} 
can be rewritten in the following scale-invariant forms
\begin{eqnarray}
P_{\Gamma}^{+}(x \vert x_0)  
&& = R_{\Gamma}^{+}(x) e^{ - \int_{x_0+\Gamma}^x dz R_{\Gamma}^+(z)  } 
= \frac{ 1 }{ \Gamma  } r^{+} \left( \frac{x}{\Gamma} \right) 
e^{ - \int_{x_0+\Gamma}^x \frac{ dz }{ \Gamma  } r^{+} \left( \frac{z}{\Gamma} \right)   } 
\nonumber \\ && 
=  \frac{ 1 }{ \Gamma  } r^{+} \left( \frac{x}{\Gamma} \right) 
e^{ - \int_{\frac{x_0}{\Gamma}+1}^{\frac{x}{\Gamma}} dw r^{+} (w)   } 
\equiv  \frac{ 1 }{ \Gamma  } 
p^{+}\left( \frac{x}{\Gamma} \bigg\vert \frac{x_0}{\Gamma} \right)
\text{ for } \ \  x \in ]x_0+\Gamma,+\infty[
\nonumber \\
P_{\Gamma}^-(x \vert x_0)  
&&= R_{\Gamma}^-(x) e^{ - \int_x^{x_0-\Gamma} dz R_{\Gamma}^-(z)  } 
=  \frac{ 1 }{ \Gamma  } r^{-} \left( \frac{x}{\Gamma} \right) e^{ - \int_x^{x_0-\Gamma} dz  \frac{ 1 }{ \Gamma  } r^{-} \left( \frac{z}{\Gamma} \right)  } 
 \nonumber \\ && 
= \frac{ 1 }{ \Gamma  } r^{-} \left( \frac{x}{\Gamma} \right) e^{ - \int_{\frac{x}{\Gamma}}^{\frac{x_0}{\Gamma}-1} dw r^{-} (w)  } \equiv  \frac{ 1 }{ \Gamma  } 
p^{-}\left( \frac{x}{\Gamma} \bigg\vert \frac{x_0}{\Gamma} \right)
\text{ for } \ \  x \in ]0,x_0-\Gamma]
  \label{P+-rescal}
\end{eqnarray}

The densities of extrema of Eq \ref{densities} with $c=0$ corresponding to the appropriate current-solution as in Eq. \ref{densitiesdriftcurrent}
\begin{eqnarray}
n^+_{\Gamma}(x) && =-\kappa 
\frac{\gamma \gamma_+ \gamma_-  x^{\gamma-1} (x-\Gamma)^{\gamma}}
{\left[\gamma_+ x^{\gamma}-  \gamma_- (x-\Gamma)^{\gamma}\right]^2} 
\ \ \text{ for } \ \ x \in ]\Gamma,+\infty[
\nonumber \\
n^-_{\Gamma}(x) && =-\kappa
\frac{\gamma \gamma_+ \gamma_- x^{\gamma-1} (x+\Gamma)^{\gamma}}
{\left[\gamma_+ (x+\Gamma)^{\gamma}-  \gamma_- x^{\gamma}\right]^2} 
\ \ \text{ for } \ \ x \in ]0,+\infty[
\label{densitiesgammaneg}
\end{eqnarray}

For a particle starting at $x \in ]0,b[$ in the internal state $\sigma$,
the exit probability $E^{-}_{[0,b]}(x,\sigma)  $ via the origin $a=0$  
depends on the behavior of the scale function $s_-(a \to 0^+)$ near the origin as follows
\begin{eqnarray}
E^{-}_{[0,b]}(x,\sigma) && =  \frac{s_+(b)-s_{\sigma}(x)}{s_+(b)-s_-(0^+)}
= \begin{cases} 
0 \ \ & \text{if } \ s_-(0^+)=-\infty \ \ \ \text{ for $\gamma \leq 0$ : the boundary $0$ is 
Never-Reached } \\
>0 & \text{if } \ s_-(0^+) > -\infty \ \ \ \text{for $\gamma > 0$ : the particle flows towards $0$ in Infinite-time}
\end{cases}
\label{Exit-0}
\end{eqnarray}
As discussed in detail for the diffusion processes in the previous work \cite{c_RGdiff},
when the boundary $0$ is never reached, 
the formulas above are the whole story, 
while when the particle flows towards $0$ in Infinite-time, one needs 
to analyze in more details the descending bonds, as discussed in the next subsection.



\subsubsection{ Case  $\gamma>0$ (Infinite-Time) and the representative example $\gamma=1$}

The normalization of the descending bond probability $P_{\Gamma}^-(x \vert x_0) $ the of Eq. \ref{P+-rescal}
on $x \in ]0,x_0-\Gamma] $
\begin{eqnarray}
\int_0^{x_0-\Gamma}  dx P_{\Gamma}^-(x \vert x_0) 
=   \left[ e^{ - \int_x^{x_0-\Gamma} dz R_{\Gamma}^-(z)  } \right]_0^{x_0-\Gamma}
= 1 - e^{ - \int_0^{x_0-\Gamma} dz R_{\Gamma}^-(z)  } \equiv 1 - W_{\Gamma}(x_0)
  \label{Pmoins0norma}
\end{eqnarray}
involves the missing weight 
\begin{eqnarray}
W_{\Gamma}(x_0) \equiv  e^{ - \int_0^{x_0-\Gamma} dz R_{\Gamma}^-(z)  } 
  \label{Wmissing}
\end{eqnarray}
that represents the probability that the particle flows towards zero 
without any further drawup $\Gamma$ (see \cite{c_RGdiff} for the discussion
concerning the equivalent case of the Geometric Brownian motion \cite{Peters,reviewGBM,c_carleman}).

To be more concrete, let us consider the special case $ \gamma=1$
where Eq. \ref{Rx+-free}
simplify into
\begin{eqnarray}
R_{\Gamma}^+(z) &&  
=  \frac{   \gamma_+ }{ z +  \gamma_- \Gamma}  
\ \ \text{ for } \ \ z \in ]\Gamma,+\infty[
\nonumber \\
R_{\Gamma}^-(z) && 
= \frac{  \gamma_- }
{z+\gamma_+ \Gamma} 
\ \ \text{ for } \ \ z \in ]0,+\infty[
\label{Rx+-free1}
\end{eqnarray}

with the corresponding bond probabilities
\begin{eqnarray}
P_{\Gamma}^{+}(x \vert x_0)  
&& = \frac{   \gamma_+ }{ x +  \gamma_- \Gamma} e^{ - \int_{x_0+\Gamma}^x dz \frac{   \gamma_+ }{ z +  \gamma_- \Gamma}  } 
=  \frac{   \gamma_+ (x_0 +  \gamma_+ \Gamma)^{\gamma_+}}
{ (x +  \gamma_- \Gamma)^{1+\gamma_+}}
\text{ for } \ \  x \in ]x_0+\Gamma,+\infty[
\nonumber \\
P_{\Gamma}^-(x \vert x_0)  
&&= R_{\Gamma}^-(x) e^{ - \int_x^{x_0-\Gamma} dz R_{\Gamma}^-(z)  } 
=   \frac{   \gamma_- (x +  \gamma_+ \Gamma)^{\gamma_- -1}}
{ (x_0 +  \gamma_- \Gamma)^{\gamma_-}}
\text{ for } \ \  x \in ]0,x_0-\Gamma]
  \label{P+-rescal1}
\end{eqnarray}
and
the missing weight of Eq. \ref{Wmissing}
\begin{eqnarray}
W_{\Gamma}(x_0) =  e^{ - \int_0^{x_0-\Gamma} dz R_{\Gamma}^-(z)  } 
= e^{ - \int_0^{x_0-\Gamma} dz \frac{  \gamma_- }
{z+\gamma_+ \Gamma}  } 
= \left( \frac{\gamma_+ \Gamma}{x_0 +  \gamma_- \Gamma} \right)^{\gamma_-}
  \label{Wmissing1}
\end{eqnarray}
while the extrema densities of Eq. \ref{densitiesgammaneg} read
\begin{eqnarray}
n^+_{\Gamma}(x) && =-\kappa 
\frac{ \gamma_+ \gamma_-   (x-\Gamma)}
{(x +  \gamma_- \Gamma)^2} 
\ \ \text{ for } \ \ x \in ]\Gamma,+\infty[
\nonumber \\
n^-_{\Gamma}(x) && =-\kappa
\frac{ \gamma_+ \gamma_-  (x+\Gamma)}
{(x +  \gamma_+ \Gamma)^2} 
\ \ \text{ for } \ \ x \in ]0,+\infty[
\label{densitiesgammaneg1}
\end{eqnarray}


\subsubsection{ Example $\gamma=-1 $ as representative example of $\gamma<0$ (Never-Reached)}

To be more concrete, let us consider the special case $ \gamma=-1$
where Eq. \ref{Rx+-free}
simplify into
\begin{eqnarray}
R_{\Gamma}^+(z) && 
= \frac{ \gamma_+(z-\Gamma)}{z (z+\gamma_+ \Gamma)} 
= \frac{ 1+\gamma_+}{ z+\gamma_+ \Gamma}  - \frac{1}{z}
\nonumber \\
R_{\Gamma}^-(z) && 
= \frac{ \gamma_-(z+\Gamma)}{z (z+\gamma_- \Gamma)} 
=  \frac{ \gamma_- -1 }{z+\gamma_- \Gamma} + \frac{1}{z}
\label{Rx+-free-1}
\end{eqnarray}
with the corresponding bond probabilities
\begin{eqnarray}
P_{\Gamma}^{+}(x \vert x_0)  
&& = \left(  \frac{ 1+\gamma_+}{ x+\gamma_+ \Gamma}  - \frac{1}{x}\right) 
e^{ - \int_{x_0+\Gamma}^x dz  \left(  \frac{ 1+\gamma_+}{ z+\gamma_+ \Gamma}  - \frac{1}{z}\right) } 
=\frac{ \gamma_+(x-\Gamma)[x_0+(\gamma_++1) \Gamma]^{1+\gamma_+} }
{(x_0+\Gamma)[x+\gamma_+ \Gamma]^{2+\gamma_+} } 
\nonumber \\
P_{\Gamma}^-(x \vert x_0)  
&&
= \left( \frac{ \gamma_- -1 }{x+\gamma_- \Gamma} + \frac{1}{x} \right) e^{ - \int_x^{x_0-\Gamma} dz \left( \frac{ \gamma_- -1 }{z+\gamma_- \Gamma} + \frac{1}{z} \right)  } 
=\frac{ \gamma_-(x+\Gamma)[x+\gamma_- \Gamma]^{\gamma_- -2} }
{(x_0-\Gamma)[x_0+(\gamma_- -1) \Gamma]^{\gamma_- -1}} 
  \label{P+-free-1}
\end{eqnarray}
while the extrema densities of Eq. \ref{densitiesgammaneg} read
\begin{eqnarray}
n^+_{\Gamma}(x) && =\kappa 
\frac{ \gamma_+ \gamma_-   (x-\Gamma)}
{\left[ x + \gamma_+ \Gamma  \right]^2} 
\ \ \text{ for } \ \ x \in ]\Gamma,+\infty[
\nonumber \\
n^-_{\Gamma}(x) && =\kappa
\frac{ \gamma_+ \gamma_-  (x+\Gamma)}
{\left[x +  \gamma_- \Gamma \right]^2} 
\ \ \text{ for } \ \ x \in ]0,+\infty[
\label{densitiesgammaneg-1}
\end{eqnarray}


\subsubsection{ Case $\gamma=0$ }

\label{subsec_gamma0}

For the special case $\gamma=0$, the scale functions satisfying Eqs \ref{derispm} and \ref{deris}
\begin{eqnarray}
 s(x) && =   1 
 \nonumber \\
 s_{+}(x) &&=   \gamma_+ \ln x
  \nonumber \\
 s_{-}(x) &&=   \gamma_+ \ln x -1
 \label{sfree0}
\end{eqnarray}
yield that
the functions of Eqs \ref{Rxplus} \ref{Rxmoins}
involve logarithms
\begin{eqnarray}
R_{\Gamma}^+(z) && = \frac{  s_+'(z)}{s_+(z)-s_-(z-\Gamma)} 
=  \frac{  \frac{\gamma_+}{z}}{\gamma_+ \ln \frac{z}{z-\Gamma} +1}
\ \ \text{ for } \ \ z \in ]\Gamma,+\infty[
\nonumber \\
R_{\Gamma}^-(z) && = \frac{  s_-'(z)}{s_+(z+\Gamma)-s_-(z)}
=  \frac{  \frac{\gamma_+}{z}}{\gamma_+ \ln \frac{ z+\Gamma}{z}  +1}
\ \ \text{ for } \ \ z \in ]0,+\infty[
\label{R+-0}
\end{eqnarray}
as the extrema densities of Eq. \ref{densitiesJ0} corresponding to vanishing-current $\kappa=0$
\begin{eqnarray}
n^+_{\Gamma}(x) && =c \frac{\frac{\gamma_+}{x} }{\left[\gamma_+ \ln \frac{x}{x-\Gamma} +1\right]^2} 
\nonumber \\
n^-_{\Gamma}(x) && =c \frac{\frac{\gamma_+}{x} }{\left[\gamma_+ \ln \frac{ x+\Gamma}{x}  +1\right]^2} 
\label{densitiesJ00}
\end{eqnarray}


\subsection{ Exponential functional of the telegraph process
corresponding to uniform rates $\gamma_{\pm}(x)=\gamma_{\pm} $ 
and linear velocities $v_{\pm}(x)=x \pm 1$ on $x \in ]1,+\infty[$  }

Let us now give an example with a normalizable steady state,
where $x(t)$ corresponds to the following exponential functional of the
telegraph process $\sigma(t)$ with the uniform rates $\gamma_{\pm}(x)=\gamma_{\pm} $
\begin{eqnarray}
x(t)  = \int_0^t dt' e^{ \int_{t'}^t dt'' \sigma(t'') }
\label{kestenint}
\end{eqnarray}
studied previously in \cite{c_diffcorre,c_largeDevRun}
while analogous exponential functionals of Brownian motion have been also much studied \cite{c_flux,c_hyperbolic,c_yor,yor}.

The dynamics of $x(t)$ defined by Eq. \ref{kestenint}
\begin{eqnarray}
\dot x(t)  = 1 +  \sigma(t) x(t)
\label{kestenderi}
\end{eqnarray}
corresponds to the run-and-tumble dynamics of Eq. \ref{runx} 
where the two velocities associated to the two states $\sigma=\pm 1$
are linear
\begin{eqnarray}
v_+(x) = x +1
\nonumber \\
v_-(x) = x-1
\label{forcelin}
\end{eqnarray}
and defined on $x \in ]1, +\infty[$

For $\gamma \equiv \gamma_+-\gamma_->0$,
the normalizable steady state $ P^*_{\pm}( x )$ 
(see \cite{c_diffcorre,c_largeDevRun} where the normalization C is given in terms of the Euler-Gamma function)
\begin{eqnarray}
P_+^*( x) &&    = C \frac{(x-1)^{\gamma_-}}{(x+1)^{\gamma_++1} }
\nonumber \\
P_-^*( x) &&    = C \frac{(x-1)^{\gamma_--1}}{(x+1)^{\gamma_+} }
\label{steadytele}
\end{eqnarray}
display the common power-law behavior for large $x$
\begin{eqnarray}
 P^*_{\pm}( x)   \opsimeq_{x \to +\infty}     \frac{  C   }
 { x^{ \gamma +1}    }
\label{steadyteleasymp}
\end{eqnarray}

 For the scale functions satisfying Eqs \ref{derispm} and \ref{deris}, 
 one can choose 
\begin{eqnarray}
 s(x) && =  \frac{ (x+1)^{\gamma_+} }{ (x-1)^{\gamma_-}}
 \nonumber \\
 s_-(x) && =  \int_2^x dz  \frac{\gamma_- (z+1)^{\gamma_+} }{ (z-1)^{\gamma_-+1}}
 = - \int_2^x dz  (z+1)^{\gamma_+} \frac{d}{dz}  (z-1)^{-\gamma_-}
 = - \left[ \frac{ (z+1)^{\gamma_+}  }{ (z-1)^{\gamma_-}}\right]_2^x 
 +  \int_2^x dz (z-1)^{-\gamma_-}
\gamma_+  (z+1)^{\gamma_+-1}  
\nonumber \\
&&= s(2)-s(x) +  \int_2^x dz 
\frac{ \gamma_+  (z+1)^{\gamma_+-1}  }{(z-1)^{\gamma_-}}
 \label{sxmtele}
\end{eqnarray}
while 
\begin{eqnarray}
 s_+(x) && =  s(x)+s_-(x) = s(2) +  \int_2^x dz 
\frac{ \gamma_+  (z+1)^{\gamma_+-1}  }{(z-1)^{\gamma_-}}
 \nonumber \\
 s_+'(x) &&= \gamma_+    \frac{ (x+1)^{\gamma_+-1} }{ (x-1)^{\gamma_-}}
 \label{sxptele}
\end{eqnarray}

Since the general expressions for the 
renormalized properties
 are then somewhat heavy,
let us focus on the behaviors for large $x$ using
\begin{eqnarray}
s_-(x-\Gamma) && \opsimeq_{x \to + \infty} s_-(x) - \Gamma s_-'(x)  + ...  
\nonumber \\
s_+(x+\Gamma) && \opsimeq_{x \to + \infty} s_+(x) + \Gamma s_+'(x)  + ...
\label{spmdevlarge}
\end{eqnarray}
Then the leading behaviors for the two functions of Eqs \ref{Rxplus} \ref{Rxmoins}
\begin{eqnarray}
R_{\Gamma}^+(x) = \frac{  s_+'(x)}{s_+(x)-s_-(x-\Gamma)} 
\opsimeq_{x \to + \infty}\frac{  s_+'(x)}{s(x)+\Gamma s_-'(x)+...} \opsimeq_{x \to + \infty}\frac{  s_+'(x)}{s(x)}\opsimeq_{x \to + \infty} \frac{\gamma_+}{x}
\nonumber \\
R_{\Gamma}^-(x) =\frac{  s_-'(x)}{s_+(x+\Gamma)-s_-(x)}
\opsimeq_{x \to + \infty} \frac{  s_-'(x)}{s(x)+\Gamma s_+'(x)+...}
\opsimeq_{x \to + \infty} \frac{  s_-'(x)}{s(x)}\opsimeq_{x \to + \infty} \frac{\gamma_-}{x}
\label{R+-large}
\end{eqnarray}
As a consequence, the probability of Eq. \ref{Pplus}
for ascending bonds decays as the power-law
\begin{eqnarray}
P_{\Gamma}^+(x \vert x_0)  
= R_{\Gamma}^+(x) e^{ - \int_{x_0+\Gamma}^x dz R_{\Gamma}^+(z)  } 
\oppropto_{x \to + \infty} \frac{\gamma_+}{x}e^{ - \int^x dz \frac{\gamma_+}{z}  } 
\oppropto_{x \to + \infty} \frac{\gamma_+}{x^{1+\gamma_+}}
  \label{Pplustele}
\end{eqnarray}
while the asymptotic behavior for the descending bonds that require that $(x_0-\Gamma)$ is even bigger than $x$
\begin{eqnarray}
P_{\Gamma}^-(x \vert x_0)  
= R_{\Gamma}^-(x) e^{ - \int_x^{x_0-\Gamma} dz R_{\Gamma}^-(z)  } 
\opsimeq_{x \to + \infty} \frac{\gamma_-}{x}e^{ - \int_x^{x_0-\Gamma} dz \frac{\gamma_-}{z}  } 
\opsimeq_{x \to + \infty} \frac{\gamma_- x^{\gamma_- -1}}{(x_0-\Gamma)^{\gamma_-}}
  \label{Pmoinstele}
\end{eqnarray}

Using again Eq. \ref{spmdevlarge}, one obtains
that the densities of extrema given by the vanishing-current expression of Eq \ref{densitiesJ0} 
\begin{eqnarray}
n^+_{\Gamma}(x) && =c \frac{s_+'(x) }{\left[s_+(x)-s_-(x-\Gamma)\right]^2} 
\opsimeq_{x \to + \infty} c \frac{s_+'(x) }{s^2(x)} 
\opsimeq_{x \to + \infty}  c\frac{\gamma_+}{x^{1+\gamma}}
\nonumber \\
n^-_{\Gamma}(x) && =c \frac{s_-'(x) }{\left[s_+(x+\Gamma)-s_-(x)\right]^2} 
\opsimeq_{x \to + \infty} c \frac{s_-'(x) }{s^2(x)} 
\opsimeq_{x \to + \infty}  c\frac{\gamma_-}{x^{1+\gamma}}
\label{densitiesJ0large}
\end{eqnarray}
display the same power-law behaviors for large $x$ as the steady state 
of Eq. \ref{steadytele} and \ref{steadyteleasymp}.


\section{ Conclusion }

\label{sec_conclusion}

In summary, the statistical properties of the renormalized trajectory at scale $\Gamma$,
where one keeps only the alternating sequence of local minima and local maxima that are separated by drawups and drawdowns of amplitude bigger than $\Gamma$
have been computed for one-dimensional run-and-tumble processes involving arbitrary velocities $v_{\pm}(x)>0$ and arbitrary switching rates $\gamma_{\pm}(x)>0$.
These explicit results generalize directly the recent work \cite{c_RGdiff} concerning diffusion processes involving arbitrary diffusion coefficients $D(x)$ and Ito forces $F_I(x)$. The main result is that the scale function $s(x)$ of diffusion processes that depend only on the ratio $\frac{F_I(x)}{D(x) } $ is replaced by a pair of scale functions $s_{\pm}(x)$ that depend only on the two ratios $\frac{\gamma_+( x ) }{v_+( x )}$ and $\frac{\gamma_-( x ) }{v_-( x )}$. This pair of scale functions $s_{\pm}(x)$ govern the exit probabilities from an interval and determine the universality classes of the renormalized trajectories: all run-and-tumble processes with the same two ratios $\frac{\gamma_+( x ) }{v_+( x )}$ and $\frac{\gamma_-( x ) }{v_-( x )}$ share the same statistics of renormalized trajectories. We have illustrated this general framework 
via simple examples : uniform symmetric and asymmetric processes on the whole line $x \in]-\infty,+\infty[$, geometric process on the half-line $x \in]0,+\infty[$, and an exponential functional of the telegraph process with a normalizable steady state.
The main differences with their diffusive counterparts can be summarized as follows:

(i) in contrast to diffusion processes, 
the initial condition of the RG at $\Gamma=0$ is well-defined for run-and-tumble processes,
since any initial stochastic trajectory is already an alternating sequence of local minima and local maxima corresponding to the switching events between the ballistic motions with velocities $v_{\pm}(x) $ in the two directions.

(ii) in the uniform case, the persistence length $\frac{v}{\gamma }$ introduces a crossover
between the ballistic regime $\Gamma \ll \frac{v}{\gamma }$ and the diffusion regime
$\Gamma \gg \frac{v}{\gamma }$. For space-dependent parameters, when the persistence length grows with $x$, as in the examples of the geometric run-and-tumble process and the exponential functional of the telegraph process, the region of large $x$ can remain ballistic at any fixed RG-scale $\Gamma$.

In the Appendices, we describe how the additional durations of the drawups and drawdowns of the renormalized trajectories can be taken into account, in order to
obtain explicit results for the time-Laplace transforms 
as a function of the Laplace variable $p>0$ (instead of $p=0$ in the main text,
where one keeps only the values of the local maxima and minima).

Further work is needed to see if the present analysis concerning run-and-tumble processes
can be generalized for other emblematic active processes.


\appendix

\section{ Distributions $\pi^{\pm}_{[a,b]} (t \vert x,\sigma) $ of the boundary-exit-times and their Laplace transforms ${\hat \pi}^{[p]\pm}_{[a,b]} ( x,\sigma) $}

\label{app_ExitTime}

\subsection{ Equations satisfied by 
the Laplace transforms ${\hat \pi}^{[p]\pm}_{[a,b]} ( x,\sigma) $
of the exit-time probabilities $\pi^{\pm}{[a,b]} ( t \vert x,\sigma)  $  }

The exit probability $E^+_{[a,b]}(x,\sigma)$ discussed in Eq. \ref{Exit+}
can be decomposed with respect to the probability distribution
$\pi^+_{[a,b]}(t \vert x,\sigma)$ of the corresponding exit time via $b$
\begin{eqnarray}
E^+_{[a,b]}(x,\sigma)= \int_0^{+\infty} dt \pi^+_{[a,b]}(t \vert x,\sigma)
  \label{ExitRxt}
\end{eqnarray}

The Laplace transform
\begin{eqnarray}
{\hat \pi}^{[p]+}_{[a,b]} ( x,\sigma) \equiv \int_0^{+\infty} dt e^{-p t}  
\pi^+_{[a,b]}(t \vert x,\sigma)
  \label{Laplace}
\end{eqnarray}
 satisfies the differential system involving the adjoint operator ${\cal L}^{\dagger} $ of Eq. \ref{runadjoint}
 \begin{eqnarray}
p   \begin{pmatrix} 
 {\hat \pi}^{[p]+}_{[a,b]} ( x,+)
 \\  {\hat \pi}^{[p]+}_{[a,b]} ( x,-) 
  \end{pmatrix}
= {\cal L}^{\dagger} 
  \begin{pmatrix} 
  {\hat \pi}^{[p]+}_{[a,b]} ( x,+)
 \\  {\hat \pi}^{[p]+}_{[a,b]} ( x,-) 
  \end{pmatrix}
  \label{LaplaceEqadjoint}
\end{eqnarray}
with the boundary conditions analog to Eq. \ref{BCExit}
\begin{eqnarray}
{\hat \pi}^{[p]+}_{[a,b]}(x=b,\sigma=+) && =1
\nonumber \\
{\hat \pi}^{[p]+}_{[a,b]}(x=a,\sigma=-) && =0
\label{BCExitb}
\end{eqnarray}

Similarly, the Laplace transform ${\hat \pi}^{[p]-}_{[a,b]} ( x,\sigma) $ for the exit time via $a$ satisfies the same Eq. \ref{LaplaceEqadjoint}
but with the different boundary conditions
\begin{eqnarray}
{\hat \pi}^{[p]-}_{[a,b]}(x=b,\sigma=+) && =0
\nonumber \\
{\hat \pi}^{[p]-}_{[a,b]}(x=a,\sigma=-) && =1
\label{BCExita}
\end{eqnarray}


\subsection{ Solution for the Laplace transforms  ${\hat \pi}^{[p]\pm}_{[a,b]} ( x,\sigma) $ }

 Let us write Eq. \ref{LaplaceEqadjoint} explicitly for the two components of the 
 vector 
 $ \begin{pmatrix} 
 g(x,+)
 \\  g(x,-) 
  \end{pmatrix}$ 
\begin{eqnarray}
v_+( x )  g'(x,+)   && = [p +   \gamma_+( x )] g(x,+) -     \gamma_+( x )g(x,-)
\nonumber \\
v_-( x )  g'(x,-)  && = \gamma_-( x )  g(x,+) -   [\gamma_-( x )+p] g(x,-)    
\label{LaplacepEq}
\end{eqnarray}
The determinant of two independent solutions 
$ \begin{pmatrix} 
 g_1^{[p]}(x,+)
 \\  g_1^{[p]}(x,-) 
  \end{pmatrix}$
  and
  $ \begin{pmatrix} 
 g_2^{[p]}(x,+)
 \\  g_2^{[p]}(x,-) 
  \end{pmatrix}$
  \begin{eqnarray}
w^{[p]}(x) \equiv g_1^{[p]}(x,+)g_2^{[p]}(x,-) -g_1^{[p]}(x,-) g_2^{[p]}(x,+)
\label{wronskian}
\end{eqnarray}
has for derivative using Eq \ref{LaplacepEq}
  \begin{eqnarray}
w^{[p]'}(x) && 
= g_1^{[p]'}(x,+)g_2^{[p]}(x,-) 
+g_1^{[p]}(x,+)g_2^{[p]'}(x,-)
-g_1^{[p]'}(x,-) g_2^{[p]}(x,+)
-g_1^{[p]}(x,-) g_2^{[p]'}(x,+)
\nonumber \\
&& = \left[  \frac{p +   \gamma_+( x )}{v_+(x)}
- \frac{p +   \gamma_-( x )}{v_-(x)} \right] w^{[p]}(x)
\label{wronskianderi}
\end{eqnarray}

As a consequence, it is useful to introduce the function
\begin{eqnarray}
 s_p(x) && \equiv  e^{\displaystyle  \int_0^x dy \left[ \frac{ p+ \gamma_+(y)    }{v_+(y) } - \frac{ p+\gamma_-(y) }{v_-( y)}  \right] }    
 = s(x) e^{\displaystyle p \int_0^x dy \left[ \frac{ 1    }{v_+(y) } - \frac{ 1 }{v_-( y)}  \right] } 
 \label{sxp}
\end{eqnarray}
that coincides with $s(x)$ of Eq. \ref{sx} for $p=0$, and to construct the ratio
that does not depend on $x$
\begin{eqnarray}
 W^{[p]} \equiv \frac {w^{[p]}(x)}{s_p(x)}
 \label{Wsxp}
\end{eqnarray}

The solution ${\hat \pi}^{[p]+}_{[a,b]}(x,\sigma)  $ satisfying the boundary conditions of Eq. \ref{BCExitb} reads
  \begin{eqnarray}
{\hat \pi}^{[p]+}_{[a,b]}(x,\sigma) = 
\frac{ g_1^{[p]}(x,\sigma)g_2^{[p]}(a,-) - g_2^{[p]}(x,\sigma)g_1^{[p]}(a,-)  }
{ g_1^{[p]}(b,+)g_2^{[p]}(a,-) - g_2^{[p]}(b,+)g_1^{[p]}(a,-)}
\label{SoluLaplacepi+}
\end{eqnarray}
while the solution ${\hat \pi}^{[p]-}_{[a,b]}(x,\sigma)  $ satisfying the boundary conditions of Eq. \ref{BCExita} is given by
  \begin{eqnarray}
{\hat \pi}^{[p]-}_{[a,b]}(x,\sigma) = 
\frac{ g_1^{[p]}(b,+)g_2^{[p]}(x,\sigma) - g_2^{[p]}(b,+)g_1^{[p]}(x,\sigma)  }
{ g_1^{[p]}(b,+)g_2^{[p]}(a,-) - g_2^{[p]}(b,+)g_1^{[p]}(a,-)}
\label{SoluLaplacepi-}
\end{eqnarray}

It is thus convenient to introduce the function of two variables
  \begin{eqnarray}
\Delta^{[p]}(x,y) \equiv g_1^{[p]}(y,+)g_2^{[p]}(x,-) - g_2^{[p]}(y,+)g_1^{[p]}(x,-)
\label{Deltap}
\end{eqnarray}
that reduces to Eq. \ref{wronskian}
at coinciding points $x=y$
  \begin{eqnarray}
\Delta^{[p]}(x,x) = w^{[p]}(x) 
\label{Deltapcoinciding}
\end{eqnarray}
in order to write the solution of Eq. \ref{SoluLaplacepi+}
for $\sigma=+$ as the ratio
  \begin{eqnarray}
{\hat \pi}^{[p]+}_{[a,b]}(x,\sigma=+) && = 
\frac{ g_1^{[p]}(x,+)g_2^{[p]}(a,-) - g_2^{[p]}(x,+)g_1^{[p]}(a,-)  }
{ g_1^{[p]}(b,+)g_2^{[p]}(a,-) - g_2^{[p]}(b,+)g_1^{[p]}(a,-)}
= \frac{\Delta^{[p]}(a,x)}{\Delta^{[p]}(a,b)}
\nonumber \\
&& =  \frac{K_+^{[p]}(a,x)}{K_+^{[p]}(a,b)}
\ \text{ with} \ K_+^{[p]}(a,x) \equiv \frac{\Delta^{[p]}(a,x)}{w^{[p]}(a)}
\ \text{ satisfying} \ K_+^{[p]}(a,x=a)=1
\label{SoluLaplacepi++}
\end{eqnarray}
and the solution of Eq. \ref{SoluLaplacepi-}
for $\sigma=-$ as the ratio
  \begin{eqnarray}
{\hat \pi}^{[p]-}_{[a,b]}(x,\sigma=-) &&= 
\frac{ g_1^{[p]}(b,+)g_2^{[p]}(x,-) - g_2^{[p]}(b,+)g_1^{[p]}(x,-)  }
{ g_1^{[p]}(b,+)g_2^{[p]}(a,-) - g_2^{[p]}(b,+)g_1^{[p]}(a,-)}
=  \frac{\Delta^{[p]}(x,b)}{\Delta^{[p]}(a,b)}
\nonumber \\
&&=  \frac{K_-^{[p]}(x,b)}{K_-^{[p]}(a,b)}
\ \text{ with} \ K_-^{[p]}(x,b) \equiv \frac{\Delta^{[p]}(x,b)}{w^{[p]}(b)}
\ \text{ satisfying} \ K_-^{[p]}(x=b,b)=1
\label{SoluLaplacepi--}
\end{eqnarray}

The introduction of the two kernels $K_{\pm}^{[p]}(.,.)$ instead of 
using only the single function $\Delta^{[p]}(.,.) $ is to have 
the boundary conditions $K_+^{[p]}(a,x=a)=1 $ and $K_-^{[p]}(x=b,b)=1 $ independent of $p$.


\subsection{ Some useful properties of $ \Delta^{[p]}(x,y)$ }

For later purposes, it is useful to consider  
the various partial derivatives of $\Delta^{[p]}(x,y) $ 
\begin{eqnarray}
 \partial_x  \Delta^{[p]}(x,y) && = 
 g_1^{[p]}(y,+)g_2^{[p]'}(x,-) - g_2^{[p]}(y,+)g_1^{[p]'}(x,-)
 \nonumber \\
 \partial_y  \Delta^{[p]}(x,y) && = 
 g_1^{[p]'}(y,+)g_2^{[p]}(x,-) - g_2^{[p]'}(y,+)g_1^{[p]}(x,-)
 \nonumber \\
  \partial_x \partial_y \Delta^{[p]}(x,y) && = 
 g_1^{[p]'}(y,+)g_2^{[p]'}(x,-) - g_2^{[p]'}(y,+)g_1^{[p]'}(x,-)
 \label{partialderiDelta}
\end{eqnarray}

Let us first focus on the combination
\begin{eqnarray}
&& \Delta^{[p]}(x,y) \partial_x \partial_y \Delta^{[p]}(x,y) 
- [\partial_x  \Delta^{[p]}(x,y) ] [ \partial_y  \Delta^{[p]}(x,y) ] 
\nonumber \\
&& =  \bigg( g_1^{[p]}(y,+) g_2^{[p]'}(y,+)- g_1^{[p]'}(y,+) g_2^{[p]}(y,+)\bigg)
\bigg( g_1^{[p]}(x,-) g_2^{[p]'}(x,-)- g_1^{[p]'}(x,-)g_2^{[p]}(x,-)\bigg)
\nonumber \\
&&=  \bigg( - \frac{  \gamma_+( y )}{v_+( y )} w^{[p]}(y) \bigg)
\bigg( - \frac{ \gamma_-( x )   }{ v_-( x )} w^{[p]}(x)\bigg)
 \label{partialderiDeltaCombi}
\end{eqnarray}
where we have used Eq. \ref{LaplacepEq} to obtain the last line.

The final result that will be needed later is the cross-derivative of the logarithm of
$ \Delta^{[p]}(x,y)$
\begin{eqnarray}
 \partial_x \partial_y \ln \Delta^{[p]}(x,y)  
&& = \frac{\Delta^{[p]}(x,y) \partial_x \partial_y \Delta^{[p]}(x,y) 
- [\partial_x  \Delta^{[p]}(x,y) ] [ \partial_y  \Delta^{[p]}(x,y) ] }{[\Delta^{[p]}(x,y) ]^2}
\nonumber \\
&& = \frac{  \gamma_+( y )}{v_+( y )} \times \frac{ \gamma_-( x )   }{ v_-( x )}
\times \frac{w^{[p]}(x)}{\Delta^{[p]}(x,y) ]} \times \frac{w^{[p]}(y)}{\Delta^{[p]}(x,y) ]} 
\nonumber \\
&& = \frac{  \gamma_+( y )}{v_+( y )} \times \frac{ \gamma_-( x )   }{ v_-( x )}
\times \frac{1}{K_+^{[p]}(x,y) ]} \times \frac{1}{K_-^{[p]}(x,y) ]} 
\label{CrossLogDeri}
\end{eqnarray}
where we have used Eqs \ref{SoluLaplacepi++}
and \ref{SoluLaplacepi--}
 to obtain the last line.
 
 Other properties that will be needed are 
 the partial derivatives of Eq. \ref{partialderiDelta}
 at coinciding points $x=y$, where one can plug Eq. \ref{LaplacepEq}
 to obtain
\begin{eqnarray}
 \partial_1  \Delta^{[p]}(x,x) && = 
 g_1^{[p]}(x,+)g_2^{[p]'}(x,-) - g_2^{[p]}(x,+)g_1^{[p]'}(x,-)
 = - \frac{\gamma_-( x )+p}{v_-(x)}   w^{[p]}(x)
 \nonumber \\
 \partial_2  \Delta^{[p]}(x,x) && = 
 g_1^{[p]'}(x,+)g_2^{[p]}(x,-) - g_2^{[p]'}(x,+)g_1^{[p]}(x,-)
 = \frac{\gamma_+( x )+p}{v_+(x)}   w^{[p]}(x)
 \label{partialderiDeltaxx}
\end{eqnarray}

As a consequence, the corresponding partial derivatives of the two kernels 
$ $ of Eq. \ref{SoluLaplacepi++} and \ref{SoluLaplacepi--}
at coinciding points simplify into
  \begin{eqnarray}
\partial_2  K_+^{[p]}(x,x) && = \frac{\partial_2 \Delta^{[p]}(x,x)}{w^{[p]}(x)}
= \frac{\gamma_+( x )+p}{v_+(x)} 
\nonumber \\
\partial_1 K_-^{[p]}(x,x) && = \frac{\partial_1\Delta^{[p]}(x,x)}{w^{[p]}(x)}
= - \frac{\gamma_-( x )+p}{v_-(x)} 
\label{SoluLaplacepi++--partialCoinciding}
\end{eqnarray}


\section{ Direct computation of the probabilities $P_{\Gamma}^{\pm}(x,t \vert x_0) $ that a renormalized bond 
at scale $\Gamma$ starting at $x_0$ ends at the position $x$ after the duration $t$ }

\subsection{ Decomposition of the duration $t=t_1+t_2$ of an ascending bond into two times }

The probability $P_{\Gamma}^+(x \vert x_0) $ of an ascending bond computed in Eq. \ref{Pplus}
of the main text can be decomposed   
\begin{eqnarray}
P_{\Gamma}^+(x \vert x_0) = \int_0^{+\infty} dt P_{\Gamma}^+(x , t\vert x_0)
  \label{Pplustdef}
\end{eqnarray}
according to its duration $t$ that 
is the sum of two contributions
\begin{eqnarray}
 && t=t_1+t_2
\nonumber \\
&& \text{$t_1$ is the time to go from the minimum $x_0$ to $(x_0+\Gamma)$ }
\nonumber \\
&& \text{$t_2$ is the time to go from $(x_0+\Gamma)$ up to the top $x$ without any drawdown $\Gamma$}
  \label{t1t2}
\end{eqnarray}
So $P_{\Gamma}^+(x , t\vert x_0) $ is given by the time-convolution 
\begin{eqnarray}
P_{\Gamma}^+(x , t\vert x_0) = \int_0^{+\infty} dt_1  \Pi^+_{\Gamma}(t_1\vert x_0)
 \int_0^{+\infty} dt_2  Q^+_{\Gamma}(x , t_2\vert x_0+\Gamma) \delta(t-(t_1+t_2))
  \label{Pplustconvolt1t2}
\end{eqnarray}
where the two probabilities $\Pi^+_{\Gamma}(t_1\vert x_0) $ and $Q^+_{\Gamma}(x , t_2\vert x_0+\Gamma) $
are analyzed in the next two subsections.


\subsection{ Probability distribution $\Pi^+_{\Gamma}(t_1\vert x_0)$ of the time $t_1$ }

The probability distribution $\Pi^+_{\Gamma}(t_1\vert x_0)$ of the time $t_1$
can be evaluated from the probability distribution $\pi^+_{[x_0,x_0+\Gamma]} (t_1 \vert x_0,+) $
of the exit-time at the right boundary
of the interval $[a=x_0,b=x_0+\Gamma ]$ when starting at $x=x_0$
and $\sigma=+$
that was analyzed in Appendix \ref{app_ExitTime}
\begin{eqnarray}
  \Pi^+_{\Gamma}(t_1\vert x_0) =  \left( \frac{\pi^+_{[x_0,x_0+\Gamma]} (t_1 \vert x_0,+)}{ \int_0^{+\infty} dt_1' \pi^+_{[x_0,x_0+\Gamma]} (t_1' \vert x_0,+) }\right)
  \label{Q1}
\end{eqnarray}
The Laplace transform ${\hat  \Pi}^{[p]+}_{\Gamma}( x_0) $
\begin{eqnarray}
{\hat  \Pi}^{[p]+}_{\Gamma}( x_0) \equiv \int_0^{+\infty} dt_1 e^{-pt_1} \Pi^+_{\Gamma}(t_1\vert x_0)
 =  \left( \frac{ {\hat \pi}^{[p]+}_{[x_0,x_0+\Gamma]} ( x_0,+)}
 { {\hat \pi}^{[p=0]+}_{[x_0,x_0+\Gamma]} ( x_0,+) }\right)
 =  \left( \frac{ {\hat \pi}^{[p]+}_{[x_0,x_0+\Gamma]} ( x_0,+)} { E^{+}_{[x_0,x_0+\Gamma]} ( x_0,+) }\right)
  \label{Q1Laplacedef}
\end{eqnarray}
thus involves the Laplace-transform ${\hat \pi}^{[p]+}_{[x_0,x_0+\Gamma]} ( x_0+\epsilon,+) $ 
computed in Eq. \ref{SoluLaplacepi++} 
\begin{eqnarray}
{\hat \pi}^{[p]+}_{[a=x_0,b=x_0+\Gamma]} (x= x_0,\sigma=+)
 && =  \frac{K_+^{[p]}(x_0,x_0)}{K_+^{[p]}(x_0,x_0+\Gamma)}
 =   \frac{1 }{ K_+^{[p]}(x_0,x_0+\Gamma) } 
  \label{pit1}
\end{eqnarray}
while the exit probability $E^{+}_{[x_0,x_0+\Gamma]} ( x_0,+) $
is given by Eq. \ref{Exit+}
\begin{eqnarray}
E^{+}_{[x_0,x_0+\Gamma]}(x_0,\sigma=+) && 
= \frac{s_+(x_0)-s_-(x_0)}{s_+(x_0+\Gamma)-s_-(x_0)}
=  \frac{s(x_0)}{s_+(x_0+\Gamma)-s_-(x_0)}
\label{Exit+needed}
\end{eqnarray}

The final result for the Laplace transform of Eq. \ref{Q1Laplacedef} is thus
\begin{eqnarray}
{\hat  \Pi}^{[p]+}_{\Gamma}( x_0) 
 = \frac{ s_+(x_0+\Gamma)-s_-(x_0) }{ s(x_0) K_+^{[p]}(x_0,x_0+\Gamma) } 
  \label{Pi1Laplace}
\end{eqnarray}


\subsection{ Probability $Q^+_{\Gamma}(x , t_2\vert x_0+\Gamma)$ 
to go from $(x_0+\Gamma)$ up to the top $x$ in time $t_2$ without any drawdown $\Gamma$ }

 The probability $Q^+_{\Gamma}(x , t_2\vert x_0+\Gamma)$ 
to go from $(x_0+\Gamma)$ up to the top $x$ in time $t_2$ without any drawdown $\Gamma$
has for Laplace transform
\begin{eqnarray}
{\hat Q}^{[p]+} _{\Gamma}(x \vert x_0+\Gamma) \equiv \int_0^{+\infty} dt_2 e^{- p t_2}
Q^+_{\Gamma}(x , t_2\vert x_0+\Gamma)
= R_{\Gamma}^+(x) e^{ - \int_{x_0+\Gamma}^x dz R_{\Gamma}^{[p]+}(z)  } 
  \label{QLplace}
\end{eqnarray}
that generalizes Eq. \ref{Pplus} of the main text concerning $p=0$ as follows :
the prefactor $R_{\Gamma}^+(x)$ is still the rate of Eq. \ref{Rxplus}
at which the running maximum at $x$ is stopped by a drawdown $\Gamma$,
while the function $R_{\Gamma}^{[p]+}(x) $ in the exponential 
appears when one evaluates the Laplace-transform 
${\hat \pi}^{[p]+}_{[a=x-\Gamma,b=x]} ( x-dx, \sigma=+) $ of the boundary-exit-time on the interval $ [a=x-\Gamma,b=x]$ when starting at $(x-dx)$ in the internal state $\sigma=+$ computed in Eq. \ref{SoluLaplacepi++}
 \begin{eqnarray}
{\hat \pi}^{[p]+}_{[a=x-\Gamma,b=x]} ( x-dx,+) = \frac{ K_+^{[p]}(x-\Gamma,x-dx)  }{ K_+^{[p]}(x-\Gamma,x) } 
=1 - dx  \frac{ \partial_2 K_+^{[p]}(x-\Gamma,x)  }{ K_+^{[p]}(x-\Gamma,x) } 
\equiv 1-dx R_{\Gamma}^{[p]+}(x)
  \label{CLLaplace+t2}
\end{eqnarray}
and reads in terms of Eq. \ref{SoluLaplacepi++}
 \begin{eqnarray}
R_{\Gamma}^{[p]+}(x) \equiv  \frac{ \partial_2 K_+^{[p]}(x-\Gamma,x)  }{ K_+^{[p]}(x-\Gamma,x) } 
= \partial_2 \ln K_+^{[p]}(x-\Gamma,x)
  \label{Rp+}
\end{eqnarray}



\subsection{ Probability $P_{\Gamma}^{+}(x,t \vert x_0)  $ of ascending bonds and its Laplace transform ${\hat P}_{\Gamma}^{[p]+}(x \vert x_0) $ }

The two previous subsections yield that the Laplace transform $ {\hat P}_{\Gamma}^{[p]+}(x \vert x_0)$ of the ascending-bond probability 
$P_{\Gamma}^{+}(x,t \vert x_0) $ of Eq. \ref{Pplustconvolt1t2}
is the product of the Laplace transforms of Eqs \ref{Pi1Laplace}
and \ref{QLplace}
\begin{eqnarray}
{\hat P}_{\Gamma}^{[p]+}(x \vert x_0)
&& \equiv \int_0^{+\infty} dt e^{- p t} P_{\Gamma}^+(x , t\vert x_0) 
 = {\hat  \Pi}^{[p]+}_{\Gamma}( x_0) {\hat Q}^{[p]+} _{\Gamma}(x \vert x_0+\Gamma)
 \nonumber \\
 && = \frac{ s_+(x_0+\Gamma)-s_-(x_0) }{ s(x_0) K_+^{[p]}(x_0,x_0+\Gamma) } 
  R_{\Gamma}^+(x) e^{ - \int_{x_0+\Gamma}^x dz R_{\Gamma}^{[p]+}(z)  } 
  \ \ \ \ \text{ for } \ \ x \in [x_0+\Gamma,+\infty[
  \label{Pplusp}
\end{eqnarray}

In particular, the probability $P_{\Gamma}^+(x=x_0+\Gamma , t\vert x_0) $ of ascending-bond about to be decimated have for Laplace transform using the explicit expression of $R_{\Gamma}^+(.)$ of Eq. \ref{Rxplus}
\begin{eqnarray}
{\hat P}_{\Gamma}^{[p]+}(x=x_0+\Gamma \vert x_0)
&&  = \frac{ s_+(x_0+\Gamma)-s_-(x_0) }{ s(x_0)K_+^{[p]}(x_0,x_0+\Gamma) } 
  R_{\Gamma}^+(x_0+\Gamma) 
  = \frac{ s_+'(x_0+\Gamma) }{ s(x_0) K_+^{[p]}(x_0,x_0+\Gamma) } 
  \label{Ppluspdeci}
\end{eqnarray}


\subsection{ Probability $P_{\Gamma}^{-}(x,t \vert x_0)  $ of descending bonds and its Laplace transform ${\hat P}_{\Gamma}^{[p]-}(x \vert x_0) $ }

For a descending bond starting at $x_0$, 
the analog of ${\hat  \Pi}^{[p]+}_{\Gamma}( x_0) $ of Eq. \ref{Pi1Laplace}
is the Laplace transform of the time $t_1$ needed to go from $x_0$ to $(x_0-\Gamma)$
\begin{eqnarray}
  {\hat  \Pi}^{[p]-}_{\Gamma}( x_0)  = \frac{ s_+(x_0)-s_-(x_0-\Gamma) }{ s(x_0)K_-^{[p]}(x_0-\Gamma,x_0) } 
   \label{Pi1Laplace-}
\end{eqnarray}
while the analog of ${\hat Q}^{[p]+} _{\Gamma}(x \vert x_0+\Gamma) $ of Eq. \ref{QLplace}
is the Laplace transform of 
the time $t_2$ needed to go from $(x_0-\Gamma)$ to $x$
\begin{eqnarray}
 {\hat Q}^{[p]-} _{\Gamma}(x \vert x_0-\Gamma)
 =  R_{\Gamma}^-(x) e^{ - \int_x^{x_0-\Gamma} dz R_{\Gamma}^{[p]-}(z)  } 
 \ \ \ \ \text{ for } \ \  x \in ]-\infty,x_0-\Gamma]
   \label{QLplace-}
\end{eqnarray}
where
 \begin{eqnarray}
R_{\Gamma}^{[p]-}(x) \equiv 
 - \frac{ \partial_1 K_-^{[p]}(x,x+\Gamma)  }{ K_-^{[p]}(x,x+\Gamma) } 
 = - \partial_1 \ln K_-^{[p]}(x,x+\Gamma)
  \label{Rp-}
\end{eqnarray}

So the Laplace transform ${\hat P}_{\Gamma}^{[p]-}(x \vert x_0)$
of the descending-bond probability $ P_{\Gamma}^{-}(x,t \vert x_0)$
that generalizes $P_{\Gamma}^-(x \vert x_0)={\hat P}_{\Gamma}^{[p=0]-}(x \vert x_0) $ of Eq. \ref{Pmoins}
of the main text
reads
\begin{eqnarray}
{\hat P}_{\Gamma}^{[p]-}(x \vert x_0)
&& \equiv \int_0^{+\infty} dt e^{- p t} P_{\Gamma}^-(x , t\vert x_0) =  {\hat  \Pi}^{[p]-}_{\Gamma}( x_0) {\hat Q}^{[p]-} _{\Gamma}(x \vert x_0-\Gamma)
 \nonumber \\
 && = \frac{ s_+(x_0)-s_-(x_0-\Gamma) }{ s(x_0)K_-^{[p]}(x_0-\Gamma,x_0) } 
  R_{\Gamma}^-(x) e^{ - \int_x^{x_0-\Gamma} dz R_{\Gamma}^{[p]-}(z)  } 
 \ \ \ \ \text{ for } \ \  x \in ]-\infty,x_0-\Gamma]
   \label{Pmoinsp}
\end{eqnarray}

In particular, the probability $P_{\Gamma}^-(x=x_0-\Gamma , t\vert x_0) $ of descending-bond about to be decimated have for Laplace transform using the explicit expression of $R_{\Gamma}^-(.)$ of Eq. \ref{Rxmoins}
\begin{eqnarray}
{\hat P}_{\Gamma}^{[p]-}(x=x_0-\Gamma \vert x_0)
 = \frac{ s_+(x_0)-s_-(x_0-\Gamma) }{ s(x_0)K_-^{[p]}(x_0-\Gamma,x_0) } 
  R_{\Gamma}^-(x_0-\Gamma) 
  =  \frac{ s_-'(x_0-\Gamma) }{ s(x_0) K_-^{[p]}(x_0-\Gamma,x_0) } 
   \label{Pmoinspdeci}
\end{eqnarray}


\subsection{ Exact initial conditions for the Laplace transform ${\hat P}_{\Gamma=0}^{[p]-}(x \vert x_0) $ at $\Gamma=0$}

At $\Gamma=0$, the time durations associated to the ascending and descending bonds 
from $x_0$ to $x$ of Eq. \ref{PIniBallisticIntervals}
involves the velocities $v_{\pm}(x)$
governing the ballistic motions on these intervals
\begin{eqnarray}
\tau_{+}(x_0,x) && = \int_{x_0}^{x} \frac{dz}{v_{+}(z)} 
\nonumber \\
\tau_{-}(x,x_0) && = \int_{x}^{x_0} \frac{dz}{v_{-}(z)} 
\label{durationsinter}
\end{eqnarray}
As a consequence, the initial Laplace transforms are given by
\begin{eqnarray}
{\hat P}^{[p]+}_{\Gamma=0}(x \vert x_0) 
= \frac{\gamma_+(x) }{v_+(x)} \ e^{\displaystyle - \int_{x_0}^{x} dz \frac{\gamma_+( z )+p}{v_+(z)}  }
\ \ \ \ \text{ for } \ \ x \in [x_0,+\infty[
\nonumber \\
{\hat P}^{[p]-}_{\Gamma=0}(x \vert x_0) 
= \frac{\gamma_-(x) }{v_-(x)} \ e^{\displaystyle - \int_{x}^{x_0} dz \frac{\gamma_-( z )+p}{v_-(z)}  }
\ \ \ \ \text{ for } \ \  x \in ]-\infty,x_0]
\label{PIniBallisticIntervalsp}
\end{eqnarray}

On the other hand, the solutions ${\hat P}_{\Gamma=0}^{[p]\pm}(x \vert x_0) $ of Eq. \ref{Pplusp}
and Eq. \ref{Pmoinsp} for $\Gamma=0$ reduce to
\begin{eqnarray}
{\hat P}_{\Gamma=0}^{[p]+}(x \vert x_0)
 && =  
  R_0^+(x) e^{ - \int_{x_0}^x dz R_0^{[p]+}(z)  } 
  \ \ \ \ \text{ for } \ \ x \in [x_0,+\infty[
\nonumber \\
{\hat P}_{\Gamma=0}^{[p]-}(x \vert x_0)
 && =  
  R_0^-(x) e^{ - \int_x^{x_0} dz R_0^{[p]-}(z)  } 
 \ \ \ \ \text{ for } \ \  x \in ]-\infty,x_0]
   \label{P+-p0}
\end{eqnarray}
where the two functions $R_{\Gamma=0}^{[p]\pm}(x) $
of Eqs \ref{Rp+}
and \ref{Rp-}
involve the partial derivatives of the functions $K_{\pm}^{[p]}$ at coinciding points
computed in Eq. \ref{SoluLaplacepi++--partialCoinciding}
\begin{eqnarray}
R_{\Gamma=0}^{[p]+}(x) = \partial_2 \ln K_+^{[p]}(x,x)
= \frac{\gamma_+( x )+p}{v_+(x)} 
\nonumber \\
R_{\Gamma}^{[p]-}(x)  = - \partial_1 \ln K_-^{[p]}(x,x)
= \frac{\gamma_-( x )+p}{v_-(x)} 
  \label{Rp+-0}
\end{eqnarray}
so that Eq. \ref{P+-p0}
is in agreement with the initial condition of Eq. \ref{PIniBallisticIntervalsp}
as it should.


\section{ RG-flows for the Laplace transforms ${\hat P}_{\Gamma}^{[p]\pm}(x \vert x_0) $ of ascending and descending bonds  } 

As explained in \cite {c_RGdiff}, the RG-flows 
for the Laplace transforms ${\hat P}_{\Gamma}^{[p]\pm}(x \vert x_0) $ of ascending and descending bonds 
are given by
\begin{eqnarray}
 \partial_{\Gamma} {\hat P}^{[p]+}_{\Gamma}(x \vert x_0 )  
&&= 
   {\hat P}^{[p]+}_{\Gamma}(x \vert x_0 ) \left[ - P_{\Gamma}^-( x-\Gamma \vert x)
 +   P_{\Gamma}^+(x_0+\Gamma \vert x_0) \right]
\nonumber \\
&&+\int_{x_0+\Gamma}^{x}d z 
{\hat P}^{[p]+}_{\Gamma}(x \vert z-\Gamma )
{\hat P}^{[p]-}_{\Gamma}( z-\Gamma \vert z) 
{\hat P}^{[p]+}_{\Gamma}(z \vert x_0 )
  \label{RGPascp}
\end{eqnarray}
and
\begin{eqnarray}
\partial_{\Gamma} {\hat P}^{[p]-}_{\Gamma}(x \vert x_0)
&& = {\hat P}^{[p]-}_{\Gamma}(x \vert x_0) 
\left[-P_{\Gamma}^+( x+\Gamma \vert x)
+ P_{\Gamma}^-( x_0-\Gamma \vert x_0)\right]
\nonumber \\
&&+\int_{x}^{x_0-\Gamma} d z 
{\hat P}^{[p]-}_{\Gamma}(x \vert z+\Gamma)
{\hat P}^{[p]+}_{\Gamma}( z+\Gamma \vert z) 
{\hat P}^{[p]-}_{\Gamma}(z \vert x_0)
  \label{RGPdesp}
\end{eqnarray}
where one can plug the Ansatz
of Eqs \ref{Pplusp}
and \ref{Pmoinsp}

\begin{eqnarray}
 {\hat P}_{\Gamma}^{[p]+}(x \vert x_0)
&& =  \frac{ s_+(x_0+\Gamma)-s_-(x_0) }{ s(x_0) K_+^{[p]}(x_0,x_0+\Gamma) } 
  R_{\Gamma}^+(x) e^{ - \int_{x_0+\Gamma}^x dz R_{\Gamma}^{[p]+}(z)  } 
  \ \ \ \ \text{ for } \ \ x \in [x_0+\Gamma,+\infty[
\nonumber \\
{\hat P}_{\Gamma}^{[p]-}(x \vert x_0)
&& =  \frac{ s_+(x_0)-s_-(x_0-\Gamma) }{ s(x_0)K_-^{[p]}(x_0-\Gamma,x_0) } 
  R_{\Gamma}^-(x) e^{ - \int_x^{x_0-\Gamma} dz R_{\Gamma}^{[p]-}(z)  } 
 \ \ \ \ \text{ for } \ \  x \in ]-\infty,x_0-\Gamma]
   \label{Pplusmoinsp}
\end{eqnarray}
to see what properties are needed to satisfy the RGflows of Eqs \ref{RGPascp}
\ref{RGPdesp}.

The log-derivatives of Eq. \ref{Pplusmoinsp}
can be simplified using Eqs \ref{Rp+} and \ref{Rp-}
as well as Eqs
\ref{Rxplusln}
\ref{Rxmoinsln}
\begin{eqnarray}
\partial_{\Gamma} \ln  {\hat P}_{\Gamma}^{[p]+}(x \vert x_0)
&& =  \frac{ s_+'(x_0+\Gamma) }{ s_+(x_0+\Gamma)-s_-(x_0) } 
- \partial_2 \ln K_+^{[p]}(x_0,x_0+\Gamma)
  +\partial_{\Gamma} \ln R_{\Gamma}^+(x)  
  + R_{\Gamma}^{[p]+}(x_0+\Gamma)
  - \int_{x_0+\Gamma}^x dz \partial_{\Gamma} R_{\Gamma}^{[p]+}(z)   
\nonumber \\  
&&  =  R_{\Gamma}^{+}(x_0+\Gamma)
  -  R_{\Gamma}^-( x-\Gamma) 
  - \int_{x_0+\Gamma}^x dz \partial_{\Gamma} R_{\Gamma}^{[p]+}(z)     
\nonumber \\
\partial_{\Gamma} \ln{\hat P}_{\Gamma}^{[p]-}(x \vert x_0)
&& =  \frac{ s_-'(x_0-\Gamma) }{ s_+(x_0)-s_-(x_0-\Gamma) } 
+ \partial_1 \ln K_-^{[p]}(x_0-\Gamma,x_0)
+\partial_{\Gamma} \ln  R_{\Gamma}^-(x) 
 + R_{\Gamma}^{[p]-}(x_0-\Gamma) 
 - \int_x^{x_0-\Gamma} dz \partial_{\Gamma} R_{\Gamma}^{[p]-}(z)   
 \nonumber \\
&& =  R_{\Gamma}^-( x_0-\Gamma) 
- R_{\Gamma}^+( x+\Gamma) 
 - \int_x^{x_0-\Gamma} dz \partial_{\Gamma} R_{\Gamma}^{[p]-}(z)   
   \label{Pplusmoinspderlog}
\end{eqnarray}

The convolution term of Eq. \ref{RGPascp} reads
\begin{eqnarray}
&& \int_{x_0+\Gamma}^{x}d z 
{\hat P}^{[p]+}_{\Gamma}(x \vert z-\Gamma )
{\hat P}^{[p]-}_{\Gamma}( z-\Gamma \vert z) 
{\hat P}^{[p]+}_{\Gamma}(z \vert x_0 )
\nonumber \\
&& =\int_{x_0+\Gamma}^{x}d z 
 \frac{ s_+(z)-s_-(z-\Gamma) }{ s(z-\Gamma) K_+^{[p]}(z-\Gamma,z) } 
  R_{\Gamma}^+(x) e^{ - \int_{z}^x dy R_{\Gamma}^{[p]+}(y)  } 
\frac{ s_-'(z-\Gamma) }{ s(z) K_-^{[p]}(z-\Gamma,z) } 
 \frac{ s_+(x_0+\Gamma)-s_-(x_0) }{ s(x_0) K_+^{[p]}(x_0,x_0+\Gamma) } 
  R_{\Gamma}^+(z) e^{ - \int_{x_0+\Gamma}^z dy R_{\Gamma}^{[p]+}(y)  } 
  \nonumber \\
&& = \frac{ s_+(x_0+\Gamma)-s_-(x_0) }{ s(x_0) K_+^{[p]}(x_0,x_0+\Gamma) }  R_{\Gamma}^+(x)  
e^{ - \int_{x_0+\Gamma}^x dy R_{\Gamma}^{[p]+}(y)  }
\int_{x_0+\Gamma}^{x}d z   
 \frac{ s_+'(z) s_-'(z-\Gamma)}{  s(z)s(z-\Gamma)   K_+^{[p]}(z-\Gamma,z)  K_-^{[p]}(z-\Gamma,z) } 
   \nonumber \\
&& = {\hat P}^{[p]+}_{\Gamma}(x \vert x_0 )
\int_{x_0+\Gamma}^{x}d z   
 \frac{ s_+'(z) s_-'(z-\Gamma)}{ s(z) s(z-\Gamma)  K_+^{[p]}(z-\Gamma,z)  K_-^{[p]}(z-\Gamma,z) } 
  \label{RGPascpconvol}
\end{eqnarray}
with the corresponding expression for 
convolution term of Eq. \ref{RGPdesp}
\begin{eqnarray}
&& \int_{x}^{x_0-\Gamma} d z 
{\hat P}^{[p]-}_{\Gamma}(x \vert z+\Gamma)
{\hat P}^{[p]+}_{\Gamma}( z+\Gamma \vert z) 
{\hat P}^{[p]-}_{\Gamma}(z \vert x_0)
\nonumber \\
&& = \int_{x}^{x_0-\Gamma} d z 
  \frac{ s_+(z+\Gamma)-s_-(z) }{ s(z+\Gamma)K_-^{[p]}(z,z+\Gamma) } 
  R_{\Gamma}^-(x) e^{ - \int_x^{z} dy R_{\Gamma}^{[p]-}(y)  } 
   \frac{ s_+'(z+\Gamma) }{ s(z) K^{[p]}_+(z,z+\Gamma) } 
    \frac{ s_+(x_0)-s_-(x_0-\Gamma) }{ s(x_0)K_-^{[p]}(x_0-\Gamma,x_0) } 
  R_{\Gamma}^-(z) e^{ - \int_z^{x_0-\Gamma} dy R_{\Gamma}^{[p]-}(y)  } 
  \nonumber \\
&& =     \frac{ s_+(x_0)-s_-(x_0-\Gamma) }{ s(x_0)K_-^{[p]}(x_0-\Gamma,x_0) } 
  R_{\Gamma}^-(x) 
e^{ - \int_x^{x_0-\Gamma} dy R_{\Gamma}^{[p]-}(y)  } 
\int_{x}^{x_0-\Gamma} d z 
     \frac{ s_-'(z) s_+'(z+\Gamma)}{s(z) s(z+\Gamma)K_-^{[p]}(z,z+\Gamma)  K^{[p]}_+(z,z+\Gamma) } 
      \nonumber \\
&& =   {\hat P}^{[p]-}_{\Gamma}(x \vert x_0)
\int_{x}^{x_0-\Gamma} d z 
     \frac{ s_-'(z) s_+'(z+\Gamma)}{s(z) s(z+\Gamma)K_-^{[p]}(z,z+\Gamma)  K^{[p]}_+(z,z+\Gamma) }
  \label{RGPdespconvol}
\end{eqnarray}

Plugging Eqs \ref{Pplusmoinspderlog}
and \ref{RGPascpconvol}
into the RG-flow of Eq. \ref{RGPascp}
yields
\begin{eqnarray}
0 && = -  \partial_{\Gamma} \ln {\hat P}^{[p]+}_{\Gamma}(x \vert x_0 )  
 - P_{\Gamma}^-( x-\Gamma \vert x)
 +   P_{\Gamma}^+(x_0+\Gamma \vert x_0) 
\nonumber \\
&&+\frac{1}{{\hat P}^{[p]+}_{\Gamma}(x \vert x_0 )} \int_{x_0+\Gamma}^{x}d z 
{\hat P}^{[p]+}_{\Gamma}(x \vert z-\Gamma )
{\hat P}^{[p]-}_{\Gamma}( z-\Gamma \vert z) 
{\hat P}^{[p]+}_{\Gamma}(z \vert x_0 )
\nonumber \\
&& = 
- R_{\Gamma}^{+}(x_0+\Gamma)
  +  R_{\Gamma}^-( x-\Gamma) 
  + \int_{x_0+\Gamma}^x dz \partial_{\Gamma} R_{\Gamma}^{[p]+}(z)   
\nonumber \\
&& - R_{\Gamma}^-( x-\Gamma )
 +   R_{\Gamma}^+(x_0+\Gamma ) 
\nonumber \\
&&+\int_{x_0+\Gamma}^{x}d z   
 \frac{ s_+'(z) s_-'(z-\Gamma)}{ s(z) s(z-\Gamma)  K_+^{[p]}(z-\Gamma,z)  K_-^{[p]}(z-\Gamma,z) } 
\nonumber \\
&& = 
 \int_{x_0+\Gamma}^x dz 
 \left[ \partial_{\Gamma} R_{\Gamma}^{[p]+}(z)   
+  
 \frac{ s_+'(z) s_-'(z-\Gamma)}{ s(z) s(z-\Gamma)  K_+^{[p]}(z-\Gamma,z)  K_-^{[p]}(z-\Gamma,z) } 
 \right]
  \label{RGPascpexpli}
\end{eqnarray}
Similarly, 
plugging Eqs \ref{Pplusmoinspderlog}
and \ref{RGPdespconvol}
into the RG-flow of Eq. \ref{RGPdesp}
yields
\begin{eqnarray}
0 && = - \partial_{\Gamma} \ln {\hat P}^{[p]-}_{\Gamma}(x \vert x_0)
 -P_{\Gamma}^+( x+\Gamma \vert x)
+ P_{\Gamma}^-( x_0-\Gamma \vert x_0)
\nonumber \\
&&+\frac{1}{{\hat P}^{[p]-}_{\Gamma}(x \vert x_0)} \int_{x}^{x_0-\Gamma} d z 
{\hat P}^{[p]-}_{\Gamma}(x \vert z+\Gamma)
{\hat P}^{[p]+}_{\Gamma}( z+\Gamma \vert z) 
{\hat P}^{[p]-}_{\Gamma}(z \vert x_0)
\nonumber \\
&& = - R_{\Gamma}^-( x_0-\Gamma) 
+ R_{\Gamma}^+( x+\Gamma) 
 + \int_x^{x_0-\Gamma} dz \partial_{\Gamma} R_{\Gamma}^{[p]-}(z)   
\nonumber \\
&& -R_{\Gamma}^+( x+\Gamma )
+ R_{\Gamma}^-( x_0-\Gamma )
\nonumber \\
&&+ \int_{x}^{x_0-\Gamma} d z 
     \frac{ s_-'(z) s_+'(z+\Gamma)}{s(z) s(z+\Gamma)K_-^{[p]}(z,z+\Gamma)  K^{[p]}_+(z,z+\Gamma) }
     \nonumber \\
&&=\int_x^{x_0-\Gamma} dz 
\left[ \partial_{\Gamma} R_{\Gamma}^{[p]-}(z)   
+     \frac{ s_-'(z) s_+'(z+\Gamma)}{s(z) s(z+\Gamma)K_-^{[p]}(z,z+\Gamma)  K^{[p]}_+(z,z+\Gamma) }
\right]
  \label{RGPdespexpli}
\end{eqnarray}

In summary, the RG-flows of Eq. \ref{RGPascpexpli}
and \ref{RGPdespexpli}
require using Eqs \ref{derispm}
\begin{eqnarray}
\partial_{\Gamma}  R_{\Gamma}^{[p]+}(z)  
  && = - \frac{ \frac{s_-'(z-\Gamma)}{s(z-\Gamma)} \times \frac{s_+'(z) }{s(z)} 
   }{   K_+^{[p]}(z-\Gamma,z)  K_-^{[p]}(z-\Gamma,z) } 
  =- \frac{ \frac{   \gamma_-( z-\Gamma ) }{v_-( z-\Gamma )} \times \frac{   \gamma_+( z ) }{v_+( z )} 
   }{   K_+^{[p]}(z-\Gamma,z)  K_-^{[p]}(z-\Gamma,z) } 
\nonumber \\
 \partial_{\Gamma} R_{\Gamma}^{[p]-}(z) 
&& = -   \frac{ \frac{s_-'(z)}{ s(z)} \times \frac{ s_+'(z+\Gamma)}{s(z+\Gamma)} }{K_-^{[p]}(z,z+\Gamma)  K^{[p]}_+(z,z+\Gamma) } 
=   -  \frac{ \frac{   \gamma_-( z ) }{v_-( z )} \times 
\frac{   \gamma_+( z+\Gamma ) }{v_+( z+\Gamma )}
 }{K_-^{[p]}(z,z+\Gamma)  K^{[p]}_+(z,z+\Gamma) } 
  \label{rgRp}
\end{eqnarray}

The derivatives with respect to $\Gamma$ of $R_{\Gamma}^{[p]\pm}(x) $ given in Eq. \ref{Rp+} 
and \ref{Rp-} 
can be evaluated 
using Eqs \ref{SoluLaplacepi++} and \ref{SoluLaplacepi--}
and the identity of Eq \ref{CrossLogDeri}
to obtain
 \begin{eqnarray}
\partial_{\Gamma} R_{\Gamma}^{[p]+}(z) 
&& =\partial_{\Gamma} \partial_2 \ln \Delta^{[p]}(z-\Gamma,z)
= - \partial_1 \partial_2 \ln \Delta^{[p]}(z-\Gamma,z)
\nonumber \\
&& = -
\frac{  \gamma_-( z-\Gamma )}{v_-( z-\Gamma )} \times \frac{ \gamma_+( z )   }{ v_+( z )}
\times \frac{1}{K_+^{[p]}(z-\Gamma,z) ]} \times \frac{1}{K_-^{[p]}(z-\Gamma,z) ]} 
  \label{Rp+deri}
\end{eqnarray}
and
 \begin{eqnarray}
\partial_{\Gamma} R_{\Gamma}^{[p]-}(z) 
&& = - \partial_{\Gamma} \partial_1 \ln \Delta^{[p]}(z,z+\Gamma)
= - \partial_2 \partial_1 \ln \Delta^{[p]}(z,z+\Gamma)
\nonumber \\
&& = - \frac{  \gamma_-( z )}{v_-( z )} \times \frac{ \gamma_+( z+\Gamma )   }{ v_+( z+\Gamma )}
\times \frac{1}{K_+^{[p]}(z,z+\Gamma) ]} \times \frac{1}{K_-^{[p]}(z,z+\Gamma) ]} 
  \label{Rp-deri}
\end{eqnarray}

In conclusion, the RG-flows of Eq. \ref{rgRp} are satisfied
as it should for consistency between the two perspectives.


\section{ Simplest example of the uniform symmetric RTP $v_{\pm}(x)=v$ and $\gamma_{\pm}(x)=\gamma $ }

Let us apply the framework of the previous Appendices to
the simplest case of the uniform symmetric RTP $v_{\pm}(x)=v$ and $\gamma_{\pm}(x)=\gamma $, when Eq. \ref{LaplacepEq}
reduces to
\begin{eqnarray}
v  g'(x,+)   && = [p +   \gamma] g(x,+) -     \gamma g(x,-)
\nonumber \\
v  g'(x,-)  && = \gamma  g(x,+) -   [\gamma+p] g(x,-)    
\label{LaplacepEqu}
\end{eqnarray}
The two independent solutions read
  \begin{eqnarray}
 \begin{pmatrix} 
 g_1^{[p]}(x,+)
 \\  g_1^{[p]}(x,-) 
  \end{pmatrix} =  \begin{pmatrix} 
 e^{x \theta_p }
 \\  r_p e^{x \theta_p } 
  \end{pmatrix}
  \ \ \ \ 
  \text{ and }
  \ \ \ 
  \begin{pmatrix} 
 g_2^{[p]}(x,+)
 \\  g_2^{[p]}(x,-) 
  \end{pmatrix} =
  \begin{pmatrix} 
 e^{- x \theta_p }
 \\  \frac{1}{r_p} e^{- x \theta_p } 
  \end{pmatrix}
\label{indepu}
\end{eqnarray}
with
\begin{eqnarray}
\theta_p && = \frac{\sqrt{ p(p+2 \gamma)} }{v}
\nonumber \\
r_p && = \frac{p+\gamma- \sqrt{ p(p+2 \gamma)} }{\gamma}
= \frac{ \gamma }{p+\gamma+ \sqrt{ p(p+2 \gamma)}  } \in ]0,1[
\label{thetaprp}
\end{eqnarray}
So Eq \ref{sxp} reduces to
 \begin{eqnarray}
s^{[p]}(x) =1
\label{spu}
\end{eqnarray}
leading to Eqs \ref{wronskian}
and \ref{Wsxp}
  \begin{eqnarray}
W^{[p]} = w^{[p]}(x) \equiv \frac{1}{r_p}  -r_p
= \frac{p+\gamma+ \sqrt{ p(p+2 \gamma)}  }{ \gamma }
-  \frac{p+\gamma- \sqrt{ p(p+2 \gamma)} }{\gamma}
= \frac{ 2 \sqrt{ p(p+2 \gamma)}  }{ \gamma }
\label{wronskianuu}
\end{eqnarray}

The function of two variables of Eq. \ref{Deltap}
  \begin{eqnarray}
\Delta^{[p]}(x,y)  && = \frac{1}{r_p} e^{ (y-x) \theta_p} - r_p e^{ - (y-x) \theta_p}
=\frac{p+\gamma+ \sqrt{ p(p+2 \gamma)}  }{ \gamma } e^{ (y-x) \theta_p} - \frac{p+\gamma- \sqrt{ p(p+2 \gamma)} }{\gamma} e^{ - (y-x) \theta_p}
\nonumber \\
&& =
\frac{ \sqrt{ p(p+2 \gamma)}  }{ \gamma } 2 \cosh[ (y-x) \theta_p] 
+\frac{p+\gamma  }{ \gamma } 2 \sinh[ (y-x) \theta_p] 
\label{Deltapu}
\end{eqnarray}
leads to
  \begin{eqnarray}
K^{[p]}_{\pm}(x,y) = \frac{\Delta^{[p]}(x,y) }{w^{[p]}} 
=  \cosh[ (y-x) \theta_p] 
+\frac{p+\gamma  }{ \sqrt{ p(p+2 \gamma)} }  \sinh[ (y-x) \theta_p]  
\label{Kpmu}
\end{eqnarray}

Using Eq. \ref{sxu},
 the Laplace transforms of Eqs \ref{Pi1Laplace} and \ref{Pi1Laplace-}
coincide 
\begin{eqnarray}
{\hat  \Pi}^{[p]\pm}_{\Gamma}( x_0) 
 = \frac{ s_+(x_0+\Gamma)-s_-(x_0) }{ s(x_0) K^{[p]}(x_0,x_0+\Gamma) } 
 = \frac{ \left[ 1+ \frac{\gamma}{v} \Gamma \right]  }
 { \cosh[ \Gamma \theta_p] 
+\frac{p+\gamma  }{ \sqrt{ p(p+2 \gamma)} }  \sinh[ \Gamma \theta_p]   } 
\equiv {\hat  \Pi}^{[p]}_{\Gamma}
  \label{Pi1Laplaceu}
\end{eqnarray}
and do not depend upon $x_0$

The two functions of
Eq. \ref{Rp+} and \ref{Rp-} coincide
 \begin{eqnarray}
R_{\Gamma}^{[p]\pm}(x) 
=\theta_p \frac{  \sinh[ \Gamma \theta_p] 
+\frac{p+\gamma  }{ \sqrt{ p(p+2 \gamma)} }  \cosh[ \Gamma \theta_p]   }
{ \cosh[ \Gamma \theta_p] 
+\frac{p+\gamma  }{ \sqrt{ p(p+2 \gamma)} }  \sinh[ \Gamma \theta_p]  } \equiv R_{\Gamma}^{[p]}
  \label{Rp+-u}
\end{eqnarray}
and do not depend on $x$
so that the Laplace transforms
of Eqs \ref{Pplusp}
and \ref{Pmoinsp}
reduce to exponentials  
\begin{eqnarray}
{\hat P}_{\Gamma}^{[p]+}(x \vert x_0)
&&  = {\hat  \Pi}^{[p]}_{\Gamma} 
  R_{\Gamma} e^{ - [x-(x_0+\Gamma)]  R_{\Gamma}^{[p]}  } 
  = \frac{  \frac{\gamma}{v}   }
 { \cosh[ \Gamma \theta_p] 
+\frac{p+\gamma  }{ \sqrt{ p(p+2 \gamma)} }  \sinh[ \Gamma \theta_p]   } 
  e^{ - [x-(x_0+\Gamma)]  R_{\Gamma}^{[p]}  }
  \ \ \ \ \text{ for } \ \ x \in [x_0+\Gamma,+\infty[
\nonumber \\
{\hat P}_{\Gamma}^{[p]-}(x \vert x_0)
 && = {\hat  \Pi}^{[p]}_{\Gamma} 
  R_{\Gamma} e^{ - [(x_0-\Gamma-x] R_{\Gamma}^{[p]}  } 
  = \frac{  \frac{\gamma}{v}   }
 { \cosh[ \Gamma \theta_p] 
+\frac{p+\gamma  }{ \sqrt{ p(p+2 \gamma)} }  \sinh[ \Gamma \theta_p]   } 
  e^{ - [(x_0-\Gamma)-x] R_{\Gamma}^{[p]}  } 
 \ \ \ \ \text{ for } \ \  x \in ]-\infty,x_0-\Gamma]
 \nonumber \\
   \label{P+-pu}
\end{eqnarray}





\begin{thebibliography}{99}



\bibitem{igloi05review}
F. Igl\'oi and C. Monthus, 
Phys. Rep. 412, 277 (2005).

\bibitem{igloiMiniReview}
F. Igl\'oi and C. Monthus, Eur. Phys. J. B (2018) 91: 290
 


\bibitem{ma79}       
       S.-K. Ma, C. Dasgupta, and C.-k. Hu,
       Phys. Rev. Lett. 43, 1434 (1979) ; \\
      C. Dasgupta and S.-K. Ma
       Phys. Rev. B 22, 1305 (1980).
       
  \bibitem{danielantiferro}
D. S. Fisher 
Phys. Rev. B 50, 3799 (1994) 

\bibitem{fisher92}
D. S. Fisher,
Phys. Rev. Lett. 69, 534 (1992) ; \\
D. S. Fisher,
Phys. Rev. B 51, 6411 (1995).

\bibitem{rgfinitesize}
D.S. Fisher and A. P. Young, 
Phys. Rev. B 58, 9131 (1998) 

\bibitem{danielreview}
D.S. Fisher, Physica A 263 (1999) 222 


\bibitem{igloi98}
F. Igl\'oi and H. Rieger.
       Phys. Rev. B {\bf 57} (1998) 11404 

\bibitem{dharyoung}
A. Dhar and A.P. Young, 
Phys. Rev. B 68, 134441 (2003) 

\bibitem{microcano}
 C. Monthus, Phys. Rev. B 69, 054431 (2004)    


\bibitem{igloi98c} F. Igl\'oi and H. Rieger, Phys. Rev. E{\bf 58}, 4238 (1998)

\bibitem{rgsinaishort}
 D. Fisher, P. Le Doussal and C. Monthus, Phys. Rev. Lett. 80  (1998) 3539


\bibitem{rgsinailong}
 D. S. Fisher, P. Le Doussal and C. Monthus,   Phys. Rev. E 59  (1999) 4795

\bibitem{rgrfim}
 D. S. Fisher, P. Le Doussal and C. Monthus,  Phys. Rev. E 64 (2001) 66107   

\bibitem{rgreadiff}
 P. Le Doussal and C. Monthus,  Phys. Rev. E 60 (1999) 1212   

\bibitem{locgolosov}
C. Monthus and  P. Le Doussal,   Phys. Rev. E 65 (2002) 66129 


\bibitem{energysinai}
 C. Monthus and P. Le Doussal, Physica A 334 (2004) 78 



\bibitem{rgtoy}
P. Le Doussal and C. Monthus,  Physica A  317  (2003) 140   

\bibitem{schehr10}
G. Schehr and P. Le Doussal,
J. Stat. Mech.  P01009 (2010).



\bibitem{Neveu}
J. Neveu and J. Pitman, 
Séminaire de Probabilités XXIII, Lecture Notes in Mathematics 1372, 239 (1989).

\bibitem{Foggionato}
A. Faggionato, Stochastic Processes and their Applications 119, 1765 (2009).



\bibitem{Taylor75}
H.M. Taylor, Annals of Prob. 3, 234 (1975).

\bibitem{Lehoczky77}
J.P. Lehoczky, Annals of Prob. 5, 601 (1977).


\bibitem{2009}
L. Pospisil, J. Vecer and O. Hadjiliadis, Stochastic Processes and their Applications 119(8), 2563 (2009).

\bibitem{Z2012}
H. Zhang and O. Hadjiliadis, Methodology and Computing in Applied Probability 14(3), 739 (2012).

\bibitem{Z2015}
H. Zhang, Advances in Applied Probability 47, 210 (2015)

\bibitem{proof}
P.J. Fitzsimmons, arXiv:2209.04695


\bibitem{Salminen}
P. Salminen and P. Vallois, 	
ESAIM: Probability and Statistics 29, 357 (2025)

\bibitem{BookZhang}
H. Zhang, "Stochastic drawdowns", World Scientific Publishing Singapore (2018)



\bibitem{c_RGdiff}
C. Monthus, arxiv:2611.04853


\bibitem{tailleur2008}
J. Tailleur and M.E. Cates, Phys. Rev. Lett. 100, 218103 (2008).

\bibitem{angelani2014}
L. Angelani, R. Di Leonardo, M. Paoluzzi, European Physical Journal E 37,59 (2014)

\bibitem{angelani2015}
L. Angelani, J. Phys. A: Math. Theor. 48, 495003 (2015)

\bibitem{evans2016}
A. B. Slowman, M. R. Evans, R. A. Blythe, Phys. Rev. Lett. 116, 218101 (2016)

\bibitem{evans2017}
 A. B. Slowman, M. R. Evans, R. A. Blythe, J Phys A: Math. Theor. 50 375601 (2017)


\bibitem{redner2018}
K. Malakar, V. Jemseena, A. Kundu, K. V. Kumar, S. Sabhapandit, S. N. Majumdar, S. Redner and A. Dhar
 J. Stat. Mech. (2018) 043215

\bibitem{satya2018}
A. Dhar, A. Kundu, S. N. Majumdar, S. Sabhapandit, G. Schehr, Phys. Rev. E 99, 032132 (2019)



\bibitem{evans2018}
M. R. Evans, S. N. Majumdar, J. Phys. A: Math. Theor. 51 475003 (2018)

\bibitem{emil2018}
E. Mallmin, R. A. Blythe, M. R. Evans, J. Stat. Mech. (2019) 013204


\bibitem{pld2019}
P. Le Doussal, S. N. Majumdar, G. Schehr, Phys. Rev. E 100, 012113 (2019)

\bibitem{harris}
M. Shreshtha, R. J. Harris, EPL 126, 40007 (2019)

\bibitem{emil2019}
E. Mallmin, R. A. Blythe, M. R. Evans, J. Phys. A: Math. Theor. 52 (2019) 425002

\bibitem{kundu2019}
P. Singh, A. Kundu, J. Stat. Mech. (2019) 083205


\bibitem{angelani}
L. Angelani, R. Garra, Modern Stochastics: Theory and Applications  6, 3 (2019)

\bibitem{emil2020}
F. Cagnetta, E. Mallmin, Phys. Rev. E 101, 022130 (2020)


\bibitem{satya2019}
U. Basu, S. N. Majumdar, A. Rosso, S. Sabhapandit, G. Schehr, J. Phys. A: Math. Theor. 53, 09LT01 (2020)

\bibitem{hartmann2020}
A. K. Hartmann, S. N. Majumdar, H. Schawe, G. Schehr, JSTAT 2020, 053401 (2020)

\bibitem{satya2020}
F. Mori, P. Le Doussal, S. N. Majumdar, G. Schehr, Phys. Rev. Lett. 124, 090603 (2020)

\bibitem{rosso2020}
T. Banerjee, S. N. Majumdar, A. Rosso, G. Schehr, Phys. Rev. E 101, 052101 (2020)

\bibitem{gregory2019}
P. Le Doussal, S. N. Majumdar, G. Schehr, EPL, 130 (2020) 40002


\bibitem{kundu2020}
Prashant Singh, Sanjib Sabhapandit, Anupam Kundu, J. Stat. Mech. (2020) 083207
 

\bibitem{pld2020}
F. Mori, P. Le Doussal, S. N. Majumdar, G. Schehr, Phys. Rev. E 102, 042133 (2020)

\bibitem{entropyprod}
N. Razin, Phys. Rev. E 102, 030103 (2020)

\bibitem{bressloff2020}
P. C Bressloff, Phys. Rev. E 102, 042135 (2020)



\bibitem{localtime}
P. Singh, A. Kundu, Phys. Rev. E 103, 042119 (2021)

\bibitem{gregory2021a}
B. De Bruyne, S. N. Majumdar, G. Schehr, J. Stat. Mech. (2021) 043211

\bibitem{boyer2021}
 G. Mercado-Vasquez, D. Boyer, Phys. Rev. E 103, 042139 (2021)

\bibitem{bressloff2021}
P. C Bressloff, arXiv:2102.10372

\bibitem{gregory2021b}
F. Mori, P. Le Doussal, S. N. Majumdar, G. Schehr, Phys. Rev. E 103, 062134 (2021)


\bibitem{c_largeDevRun}
C. Monthus, J. Stat. Mech. (2021) 083212



\bibitem{Mori}
F. Mori, G. Gradenigo, S. N. Majumdar, J. Stat. Mech. 103208 (2021)


\bibitem{Kundu}
P. Singh, A. Kundu, S. N. Majumdar, H. Schawe, J. Phys. A: Math. Theor. 55 225001 (2022)

\bibitem{Naftali}
N. R. Smith, P. Le Doussal, S. N. Majumdar, G. Schehr, Phys. Rev. E 106, 054133 (2022)

\bibitem{FvW}
T. Arnoulx de Pirey and F. van Wijland,
JSTAT 093202 (2023)

\bibitem{mathis24}
M Gu\'eneau, S. N. Majumdar, G. Schehr, EPL 145 61002 (2024)

\bibitem{Leo1}
M. Gu\'eneau and L. Touzo, J. Phys. A: Math. Theor. 57 225005 (2024)

\bibitem{Leo2}
M. Gu\'eneau and L. Touzo, J. Stat. Mech. (2024) 083208


\bibitem{mathis25}
M. Gu\'eneau, S. N. Majumdar, G. Schehr, Phys. Rev. E 111, 014144 (2025)

\bibitem{bressloff25}
P. C. Bressloff, Proc. R. Soc. A 481 (2317), 20240815 (2025)


\bibitem{Basu}
U. Basu, S. N. Majumdar, A. Rosso, APS Open Sci. 1, 000021 (2026)

\bibitem{mathis26}
M. Gu\'eneau, S. N. Majumdar, G. Schehr, Phys. Rev. E 114, 014144 (2026)




\bibitem{BookKarlin}
S. Karlin and H. Taylor, 
"A Second Course in Stochastic Processes", Academic Press, New York (1981).

\bibitem{BookYor}
D. Revuz and M. Yor, "Continuous martingales and Brownian motion",
Springer Verlag Berlin Heidelberg (1991)


\bibitem{HandBook}
A. N. Borodin and P. Salminen
"Handbook of Brownian Motion - Facts and Formulae"
Birkhäuser Verlag, Springer Basel (2002)


\bibitem{BookBorodin}
Andrei N. Borodin
"Stochastic Processes", Birkh\"auser
Springer International Publishing 
 Switzerland  (2017)
 
  

\bibitem{gardiner}
 C. W. Gardiner, “ Handbook of Stochastic Methods: for Physics, Chemistry and the Natural Sciences” (Springer Series
in Synergetics), Berlin (1985).

\bibitem{vankampen}
N.G. Van Kampen, “Stochastic processes in physics and chemistry”, Elsevier Amsterdam (1992).

\bibitem{risken}
 H. Risken, “The Fokker-Planck equation : methods of solutions and applications”, Springer Verlag Berlin (1989).


  
 
 \bibitem{glauber}
 R.J. Glauber, J. Math. Phys. 4, 294 (1963).
 
  \bibitem{Felderhof}
 B.U. Felderhof, Rep. Math. Phys. 1, 215 (1970); Rev. Math. Phys. 2, 151 (1971).
 
\bibitem{referee}
 C. F. Polnaszek, J. H. Freed,  J. Chem. Phys. 58, 3185 (1973).

\bibitem{siggia}
 E. D. Siggia, Phys. Rev. B 16, 2319 (1977).

\bibitem{kimball}
J. C. Kimball, J. Stat. Phys. 21, 289 (1979).

\bibitem{peschel}
 I. Peschel and V. J. Emery, Z. Phys. B 43, 241 (1981)
 
\bibitem{jpb_antoine}
 J.P. Bouchaud and A. Georges, Phys. Rep. 195, 127 (1990).
 
 
\bibitem{texier}
 C. Texier and C. Hagendorf, Europhys. Lett. 86 (2009) 37011.
 
\bibitem{us_eigenvaluemethod}
 C. Monthus and T. Garel, J. Stat. Mech. P12017 (2009).
 
\bibitem{Castelnovo}
 C. Castelnovo, C. Chamon and D. Sherrington, Phys. Rev. B 81, 184303 (2010).
 
 \bibitem{c_lyapunov}
C. Monthus, J. Stat. Mech. (2021) 033303.

\bibitem{us_gyrator}
A. Mazzolo and C. Monthus, Phys. Rev. E 107, 014101 (2023).

\bibitem{us_kemeny}
A. Mazzolo and C. Monthus, J. Stat. Mech. (2023) 063204.

 
\bibitem{c_pearson}
C. Monthus, J. Stat. Mech. (2023) 083204.
 
 \bibitem{c_boundarydriven}
C. Monthus,  J. Stat. Mech. (2023) 063206.



 \bibitem{c_largedevDBsusy}
 C. Monthus, J. Stat. Mech. (2024) 073203.


 \bibitem{c_susyFokkerPlanck}
C. Monthus, J. Phys. A: Math. Theor. 59 (2026) 105001


\bibitem{Peters}
O. Peters and W. Klein, Phys. Rev. Lett. 110, 100603 (2013).

\bibitem{reviewGBM}
V. Stojkoski, T. Sandev, L. Basnarkov, L. Kocarev, R. Metzler, Entropy  22, 1432 (2020)

\bibitem{c_carleman}
C. Monthus, J. Stat. Mech. (2026) 033204







\bibitem{c_diffcorre}
C. Monthus, Phys. Rev. E 52, 2569 (1995)

\bibitem{c_flux}
C. Monthus and A. Comtet, J. Phys. I France 4, 635 (1994).

\bibitem{c_hyperbolic}
A. Comtet and C. Monthus, J. Phys. A: Math. Gen. 29, 1331 (1996)

\bibitem{c_yor}
A. Comtet, C. Monthus, M. Yor, J. Appl. Proba. 35, 255 (1998).

\bibitem{yor}
M. Yor, "Exponential Functionals of Brownian Motion and Related Processes",
Springer-Verlag Berlin Heidelberg 2001




\end{thebibliography}
\end{document}